# Distribution Statistics and Random Matrix Formalism of Multicarrier Continuous-Variable Quantum Key Distribution

Laszlo Gyongyosi

[1] Quantum Technologies Laboratory, Department of Telecommunications
*Budapest University of Technology and Economics*
2 Magyar tudosok krt, Budapest, *H*-1117, Hungary
[2] MTA-BME Information Systems Research Group
*Hungarian Academy of Sciences*
7 Nador st., Budapest, *H*-1051, Hungary

gyongyosi@hit.bme.hu

**Abstract**

We propose a combined mathematical framework of order statistics and random matrix theory for multicarrier continuous-variable (CV) quantum key distribution (QKD). In a multicarrier CVQKD scheme, the information is granulated into Gaussian subcarrier CVs, and the physical Gaussian link is divided into Gaussian sub-channels. The sub-channels are dedicated to the conveying of the subcarrier CVs. The distribution statistics analysis covers the study of the distribution of the sub-channel transmittance coefficients in the presence of a Gaussian noise and the utilization of the moment generation function (MGF) in the error analysis. We reveal the mathematical formalism of sub-channel selection and formulation of the transmittance coefficients, and show a reduced complexity progressive sub-channel scanning method. We define a random matrix formalism for multicarrier CVQKD to evaluate the statistical properties of the information flowing process. Using random matrix theory, we express the achievable secret key rates and study the efficiency of the AMQD-MQA (adaptive multicarrier quadrature division–multiuser quadrature allocation) multiple-access multicarrier CVQKD. The proposed combined framework is particularly convenient for the characterization of the physical processes of experimental multicarrier CVQKD.

# 1 Introduction

The continuous-variable quantum key distribution (CVQKD) protocols allow the establishment of an unconditional secure communication over standard, currently established telecommunication networks [10–22]. In comparison to discrete-variable (DV) QKD protocols, CVQKD does not require single-photon sources and detectors and can be implemented by standard optical telecommunication devices [1], [9–26], [30–37]. In a CVQKD system, the information is carried by a continuous-variable quantum state that is defined in the phase space via the position and momentum quadratures. Since a Gaussian modulation is a reasonable modulation technique in an experiment, these CV quantum states have a Gaussian random distribution. Precisely, the presence of an eavesdropper on the quantum channel adds a white Gaussian noise into the transmission because the optimal attack against a CVQKD protocol is provably also Gaussian. Besides the attractive properties of CVQKD, the relevant performance attributes, such as secret key rates and transmission distances, still require significant improvements. For this purpose, the multicarrier CVQKD has been recently introduced through the adaptive quadrature division modulation (AMQD) scheme [2]. The multicarrier CVQKD injects several additional degrees of freedom onto the transmission which is not available for a standard, single-carrier CVQKD setting. In particular, these extra benefits and resources not just allow the realization of higher secret key rates and higher amount of tolerable losses with unconditional security but also make possible the introduction and defining of several new phenomena for CVQKD, such as singular layer transmission [4], enhanced security thresholds [5], multidimensional manifold extraction [6], and the characterization of the subcarrier domain [7]. The benefits of multicarrier CVQKD has also been proposed for multiple-access multicarrier CVQKD via the AMQD-MQA (multiuser quadrature allocation) [3]. An adaptive quadrature detection technique has also been defined for multicarrier CVQKD, which uses a channel transmittance estimation to decode the continuous variables [8].

In this work, we particularly focus on the characterization of the transmittance coefficients of the sub-channels, through the statistical analysis of their distribution. We also define a random matrix formalism to describe the process of information flow via the Gaussian subcarrier CVs. We develop a combined framework that utilizes and integrates the results of distribution statistics and random matrix theory. The proposed combined framework provides a tool to characterize the physical distribution and transmission processes of information flowing in experimental multicarrier CVQKD scenarios.

In the first part, we provide a *distribution statistics* framework for multicarrier CVQKD. The distribution statistics of multicarrier CVQKD utilizes the results of order statistics [28–29], which is an important subfield of statistical theory, with several applications—from mathematical statistics and engineering to the analysis of traditional communication systems. We reveal the statistical properties of the multicarrier transmission and define the statistical operators of sub-channel ordering, selection, and formulation of the single-carrier level transmittance coefficients. We also define the conditions that are required for the simultaneous achievement of a maximal secret key rate and an unconditional security at diverse channel parameters. The proposed distribution statistics analysis covers the study of the distribution of the sub-channel transmittance coefficient in the presence of a Gaussian noise and the utilization of the moment generation function (MGF) in the error analysis, such as the distribution of the received Gaussian subcarriers.

Random matrix theory represents a useful mathematical tool with a widespread application—from physics, statistics, to engineering problems and communication theory [27]. The popularity of random matrix theory is rooted in the fact that several problems can be directly interpreted and solved via the mathematical framework of random matrix formalism. In the second part, using the results of distributions statics, we define a *random matrix formalism* for multicarrier CVQKD. Using the framework provided by random matrix formalism, we characterize the statistical properties of the information transmission process, derive the achievable secret key rates, and study the multiuser efficiency of the AMQD-MQA multiple-access multicarrier CVQKD scheme.

This paper is organized as follows. In Section 2, some preliminary findings are summarized. Section 3 provides the distribution statistics of multicarrier CVQKD. Section 4 discusses the random matrix formalism of multiple-access multicarrier CVQKD via the analysis of AMQD-MQA. Finally, Section 5 concludes the results. Supplementary information is included in the Appendix.

## 2 Preliminaries

In Section 2, we briefly summarize the notations and basic terms. For further information, see the detailed descriptions of [2–8].

### 2.1 Basic Terms and Definitions

#### 2.1.1 Multicarrier CVQKD

The following description assumes a single user, and the use of $n$ Gaussian sub-channels $\mathcal{N}_i$ for the transmission of the subcarriers, from which only $l$ sub-channels will carry valuable information.

In the single-carrier modulation scheme, the $j$-th input single-carrier state $\left|\varphi_j\right\rangle = \left|x_j + \mathrm{i}p_j\right\rangle$ is a Gaussian state in the phase space $\mathcal{S}$, with i.i.d. Gaussian random position and momentum quadratures $x_j \in \mathbb{N}\left(0, \sigma_{\omega_0}^2\right)$, $p_j \in \mathbb{N}\left(0, \sigma_{\omega_0}^2\right)$, where $\sigma_{\omega_0}^2$ is the modulation variance of the quadratures. In the multicarrier scenario, the information is carried by Gaussian subcarrier CVs, $\left|\phi_i\right\rangle = \left|x_i + \mathrm{i}p_i\right\rangle$, $x_i \in \mathbb{N}\left(0, \sigma_\omega^2\right)$, $p_i \in \mathbb{N}\left(0, \sigma_\omega^2\right)$, where $\sigma_\omega^2$ is the modulation variance of the subcarrier quadratures, which are transmitted through a noisy Gaussian sub-channel $\mathcal{N}_i$. Precisely, each $\mathcal{N}_i$ Gaussian sub-channel is dedicated for the transmission of one Gaussian subcarrier CV from the $n$ subcarrier CVs. (*Note*: index $i$ refers to a subcarrier CV, index $j$ to a single-carrier CV, respectively.)

The single-carrier CV state $\left|\varphi_j\right\rangle$ in the phase space $\mathcal{S}$ can be modeled as a zero-mean, circular symmetric complex Gaussian random variable $z_j \in \mathcal{CN}\left(0, \sigma^2_{\omega_{z_j}}\right)$, with a variance

$$\sigma^2_{\omega_{z_j}} = \mathbb{E}\left[\left|z_j\right|^2\right] = 2\sigma^2_{\omega_0}, \tag{1}$$

and with i.i.d. real and imaginary zero-mean Gaussian random components

$$\mathrm{Re}\left(z_j\right) \in \mathbb{N}\left(0, \sigma^2_{\omega_0}\right), \; \mathrm{Im}\left(z_j\right) \in \mathbb{N}\left(0, \sigma^2_{\omega_0}\right). \tag{2}$$

In the multicarrier CVQKD scenario, let $n$ be the number of Alice's input single-carrier Gaussian states. Precisely, the $n$ input coherent states are modeled by an $n$-dimensional, zero-mean, circular symmetric complex random Gaussian vector

$$\mathbf{z} = \mathbf{x} + \mathrm{i}\mathbf{p} = \left(z_0, \ldots, z_{n-1}\right)^T \in \mathcal{CN}\left(0, \mathbf{K_z}\right), \tag{3}$$

where each $z_j$ is a zero-mean, circular symmetric complex Gaussian random variable

$$z_j \in \mathcal{CN}\left(0, \sigma^2_{\omega_{z_j}}\right), \; z_j = x_j + \mathrm{i}p_j. \tag{4}$$

In the first step of AMQD, Alice applies the inverse FFT (fast Fourier transform) operation to vector $\mathbf{z}$ (see (3)), which results in an $n$-dimensional zero-mean, circular symmetric complex Gaussian random vector $\mathbf{d}$, $\mathbf{d} \in \mathcal{CN}\left(0, \mathbf{K_d}\right)$, $\mathbf{d} = \left(d_0, \ldots, d_{n-1}\right)^T$, precisely as

$$\mathbf{d} = F^{-1}\left(\mathbf{z}\right) = e^{\frac{\mathbf{d}^T \mathbf{A}\mathbf{A}^T \mathbf{d}}{2}} = e^{\frac{\sigma^2_{\omega_0}\left(d_0^2 + \ldots + d_{n-1}^2\right)}{2}}, \tag{5}$$

where

$$d_i = x_{d_i} + \mathrm{i}p_{d_i}, \; d_i \in \mathcal{CN}\left(0, \sigma^2_{d_i}\right), \tag{6}$$

where $\sigma^2_{\omega_{d_i}} = \mathbb{E}\left[\left|d_i\right|^2\right] = 2\sigma^2_{\omega}$, thus the position and momentum quadratures of $\left|\phi_i\right\rangle$ are i.i.d. Gaussian random variables with a constant variance $\sigma^2_{\omega}$ for all $\mathcal{N}_i, i = 0, \ldots, l-1$ sub-channels:

$$\mathrm{Re}\left(d_i\right) = x_{d_i} \in \mathbb{N}\left(0, \sigma^2_{\omega}\right), \; \mathrm{Im}\left(d_i\right) = p_{d_i} \in \mathbb{N}\left(0, \sigma^2_{\omega}\right), \tag{7}$$

where $\mathbf{K_d} = \mathbb{E}\left[\mathbf{dd}^\dagger\right]$, $\mathbb{E}\left[\mathbf{d}\right] = \mathbb{E}\left[e^{\mathrm{i}\gamma}\mathbf{d}\right] = \mathbb{E}e^{\mathrm{i}\gamma}\left[\mathbf{d}\right]$, and $\mathbb{E}\left[\mathbf{dd}^T\right] = \mathbb{E}\left[e^{\mathrm{i}\gamma}\mathbf{d}\left(e^{\mathrm{i}\gamma}\mathbf{d}\right)^T\right] = \mathbb{E}e^{\mathrm{i}2\gamma}\left[\mathbf{dd}^T\right]$ for any $\gamma \in \left[0, 2\pi\right]$.

The $\mathbf{T}\left(\mathcal{N}\right)$ transmittance vector of $\mathcal{N}$ in the multicarrier transmission is

$$\mathbf{T}\left(\mathcal{N}\right) = \left[T_0\left(\mathcal{N}_0\right), \ldots, T_{n-1}\left(\mathcal{N}_{n-1}\right)\right]^T \in \mathcal{C}^n, \tag{8}$$

where

$$T_i\left(\mathcal{N}_i\right) = \mathrm{Re}\left(T_i\left(\mathcal{N}_i\right)\right) + \mathrm{i}\,\mathrm{Im}\left(T_i\left(\mathcal{N}_i\right)\right) \in \mathcal{C}, \tag{9}$$

is a complex variable, which quantifies the position and momentum quadrature transmission (i.e., gain) of the $i$-th Gaussian sub-channel $\mathcal{N}_i$, in the phase space $\mathcal{S}$, with real and imaginary parts

$$0 \leq \operatorname{Re} T_i(\mathcal{N}_i) \leq 1/\sqrt{2}, \text{ and } 0 \leq \operatorname{Im} T_i(\mathcal{N}_i) \leq 1/\sqrt{2}. \tag{10}$$

Particularly, the $T_i(\mathcal{N}_i)$ variable has the squared magnitude of

$$\left|T_i(\mathcal{N}_i)\right|^2 = \operatorname{Re} T_i(\mathcal{N}_i)^2 + \operatorname{Im} T_i(\mathcal{N}_i)^2 \in \mathbb{R}, \tag{11}$$

where

$$\operatorname{Re} T_i(\mathcal{N}_i) = \operatorname{Im} T_i(\mathcal{N}_i). \tag{12}$$

The Fourier-transformed transmittance of the $i$-th sub-channel $\mathcal{N}_i$ (resulted from CVQFT operation at Bob) is denoted by

$$\left|F(T_i(\mathcal{N}_i))\right|^2. \tag{13}$$

The $n$-dimensional zero-mean, circular symmetric complex Gaussian noise vector $\Delta \in \mathcal{CN}(0, \sigma_\Delta^2)_n$, of the quantum channel $\mathcal{N}$, is evaluated as

$$\Delta = (\Delta_0, \ldots, \Delta_{n-1})^T \in \mathcal{CN}(0, \mathbf{K}_\Delta), \tag{14}$$

where

$$\mathbf{K}_\Delta = \mathbb{E}[\Delta \Delta^\dagger], \tag{15}$$

with independent, zero-mean Gaussian random components

$$\Delta_{x_i} \in \mathbb{N}(0, \sigma_{\mathcal{N}_i}^2), \text{ and } \Delta_{p_i} \in \mathbb{N}(0, \sigma_{\mathcal{N}_i}^2), \tag{16}$$

with variance $\sigma_{\mathcal{N}_i}^2$, for each $\Delta_i$ of a Gaussian sub-channel $\mathcal{N}_i$, which identifies the Gaussian noise of the $i$-th sub-channel $\mathcal{N}_i$ on the quadrature components $x_i, p_i$ in the phase space $\mathcal{S}$. Thus $F(\Delta) \in \mathcal{CN}(0, \sigma_{\Delta_i}^2)$, where

$$\sigma_{\Delta_i}^2 = 2\sigma_{\mathcal{N}_i}^2. \tag{17}$$

The CVQFT-transformed noise vector can be rewritten as

$$F(\Delta) = (F(\Delta_0), \ldots, F(\Delta_{n-1}))^T, \tag{18}$$

with independent components $F(\Delta_{x_i}) \in \mathbb{N}(0, \sigma_{\mathcal{N}_i}^2)$ and $F(\Delta_{p_i}) \in \mathbb{N}(0, \sigma_{\mathcal{N}_i}^2)$ on the quadratures, for each $F(\Delta_i)$. Precisely, it also defines an $n$-dimensional zero-mean, circular symmetric complex Gaussian random vector $F(\Delta) \in \mathcal{CN}(0, \mathbf{K}_{F(\Delta)})$ with a covariance matrix

$$\mathbf{K}_{F(\Delta)} = \mathbb{E}\left[F(\Delta)F(\Delta)^\dagger\right]. \tag{19}$$

The complex $A_j(\mathcal{N}_j) \in \mathbb{C}$ single-carrier channel coefficient is derived from the $l$ Gaussian sub-channel coefficients as

$$A_j(\mathcal{N}_j) = \tfrac{1}{l}\sum_{i=0}^{l-1} F(T_i(\mathcal{N}_i)). \tag{20}$$

# 3 Distribution Statistics for Multicarrier CVQKD

First we summarize some preliminary findings from order statistics from [29], then evaluate the theorems and proofs. Note the proofs throughout Section 3 follow the notations of [29].

## 3.1 Moment-Generating Function

The M MGF-function (moment-generating function) [29] of a nonnegative random variable $x$, $x \geq 0$ is

$$\mathrm{M}_x(c) = \int_0^\infty P(x) e^{cx} dx, \tag{21}$$

where $c$ is a complex dummy variable, $P(\cdot)$ is the PDF (probability density function) function, and

$$\mathbb{E}\left[x^n\right] = \frac{d^n}{dc^n} \mathrm{M}_x(c)\Big|_{c=0}. \tag{22}$$

It can be shown that

$$\mathrm{M}_x(-c) = \mathcal{L}\{P(x)\}, \tag{23}$$

where $\mathcal{L}$ is the Laplace transform.

Particularly, using the M-function, the $Q(\cdot)$ Gaussian tail function of $x > 0$ can be expressed precisely as

$$Q(x) = \frac{1}{\pi} \int_0^{\pi/2} e^{\left(\frac{-x^2}{2\sin^2\phi}\right)} d\phi, \tag{24}$$

where $\phi \in [0, 2\pi]$ and

$$Q^2(x) = \frac{1}{\pi} \int_0^{\pi/4} e^{\left(\frac{-x^2}{2\sin^2\phi}\right)} d\phi. \tag{25}$$

Assuming an error rate $R_{err}(x)$

$$R_{err}(x) = a Q\left(\sqrt{bx}\right), \tag{26}$$

where $a$ and $b$ are constants, the $\widetilde{R_{err}}$ average error rate is yielded as

$$\widetilde{R_{err}} = \int_0^\infty a Q\left(\sqrt{b\mathrm{SNR}}\right) P_{\mathrm{SNR}}(x) dx, \tag{27}$$

where $P_{\mathrm{SNR}}(\cdot)$ is the PDF of the SNR (signal to noise ratio), which can be rewritten as

$$\widetilde{R_{err}} = \frac{a}{\pi} \int_0^{\pi/2} \mathrm{M}_{\mathrm{SNR}}\left(\frac{-b}{2\sin^2\phi}\right) d\phi, \tag{28}$$

where $\mathrm{M}_{\mathrm{SNR}}(\cdot)$ is the M-function of the SNR.

## 3.2 Sub-channel Distribution Statistics

### 3.2.1 Marginal, Joint and Conditional Distributions of the Sub-channels

Let the normalized independent, *ordered* transmittance coefficients of the $l$ sub-channels be as

$$\tfrac{1}{l}\left|F\left(T_i\left(\mathcal{N}_i\right)\right)\right|^2, \ i=0,\dots,l-1. \tag{29}$$

given in a descending order such that

$$\tfrac{1}{l}\left|F\left(T_0\left(\mathcal{N}_0\right)\right)\right|^2 \geq \tfrac{1}{l}\left|F\left(T_1\left(\mathcal{N}_1\right)\right)\right|^2 \geq \dots \geq \tfrac{1}{l}\left|F\left(T_{l-1}\left(\mathcal{N}_{l-1}\right)\right)\right|^2. \tag{30}$$

The $\tfrac{1}{l}\left|F\left(\widehat{T}_i\left(\mathcal{N}_i\right)\right)\right|^2$ *unordered* variables are identically distributed with common PDF and CDF (cumulative distribution function), $P_c(x)$ and $f_c(x)$, respectively [29].

The $P_c\left(x_0,\dots,x_{l-1}\right)$ common joint PDF of the unordered variables can be expressed as

$$P_c\left(x_0,\dots,x_{l-1}\right) = \prod_{i=0}^{l-1} P_c\left(x_i\right). \tag{31}$$

Without loss of generality, the ordered variables are not identically distributed and have PDF and CDF functions

$$P_{\tfrac{1}{l}\left|F(T_i(\mathcal{N}_i))\right|^2}(x) \text{ and } f_{\tfrac{1}{l}\left|F(T_i(\mathcal{N}_i))\right|^2}(x). \tag{32}$$

Then let $P_{\tfrac{1}{l}\left|F(T_i(\mathcal{N}_i))\right|^2}(x)$ the PDF function of the $i$-th ordered sub-channel coefficient, as

$$P_{\tfrac{1}{l}\left|F(T_i(\mathcal{N}_i))\right|^2}(x) = \tfrac{l!}{(l-i)!(i-1)!} f_c(x)^{l-1}\left(1-f_c(x)\right)^{l-1} P_c(x), \tag{33}$$

with

$$P_{\tfrac{1}{l}\left|F(T_0(\mathcal{N}_0))\right|^2}(x) = l f_c(x)^{l-1} P_c(x), \tag{34}$$

and

$$P_{\tfrac{1}{l}\left|F(T_{l-1}(\mathcal{N}_{l-1}))\right|^2}(x) = l\left(1-f_c(x)\right)^{l-1} P_c(x). \tag{35}$$

The $P_{\tfrac{1}{l}\left|F(T_k(\mathcal{N}_k))\right|^2,\tfrac{1}{l}\left|F(T_m(\mathcal{N}_m))\right|^2}(x,y)$ joint PDF of $\tfrac{1}{l}\left|F\left(T_k\left(\mathcal{N}_k\right)\right)\right|^2, \tfrac{1}{l}\left|F\left(T_m\left(\mathcal{N}_m\right)\right)\right|^2, m>k$ is as

$$\begin{aligned}
&P_{\tfrac{1}{l}\left|F(T_k(\mathcal{N}_k))\right|^2,\tfrac{1}{l}\left|F(T_m(\mathcal{N}_m))\right|^2}(x,y) \\
&= \tfrac{l!}{(k-1)!(m-k-1)!(l-m)!}\left(1-f_c(x)\right)^{k-1} P_c(x) \\
&\quad \cdot \left(f_c(x)-f_c(y)\right)^{m-k-1} P_c(y)\left(f_c(y)\right)^{l-m}.
\end{aligned} \tag{36}$$

Since the ordered variables are not identically distributed, the $P_c\left(x_0,\dots,x_{l-1}\right)$ common joint PDF of (31) can be rewritten as a joint PDF function

$$\begin{aligned}
&P_{\tfrac{1}{l}\left|F(T_0(\mathcal{N}_0))\right|^2,\dots,\tfrac{1}{l}\left|F(T_{l-1}(\mathcal{N}_{l-1}))\right|^2}\left(x_0,\dots,x_{l-1}\right) \\
&= l!\prod_{i=0}^{l-1} P_c(x_i), \ x_0 \geq x_1 \geq \dots \geq x_{l-1}.
\end{aligned} \tag{37}$$

Specifically, assuming that there are $l - i$ available sub-channels, the $m$-th ordered sub-channel coefficient is referred to as

$$\frac{1}{l-i}\left|F\left(T'_m\left(\mathcal{N}_m\right)\right)\right|^2, \ m = 0,\ldots,(l-i)-1. \tag{38}$$

Using the variable of (38), and further assuming that

$$\frac{1}{l}\left|F\left(T_i\left(\mathcal{N}_i\right)\right)\right|^2 = y, \tag{39}$$

the $P_{\frac{1}{l}|F(T_m(\mathcal{N}_m))|^2 \Big| \frac{1}{l}|F(T_i(\mathcal{N}_i))|^2 = y}(x)$ conditional PDF can be evaluated as

$$\begin{aligned}
P_{\frac{1}{l}|F(T_m(\mathcal{N}_m))|^2 \Big| \frac{1}{l}|F(T_i(\mathcal{N}_i))|^2 = y}(x) &= \frac{P_{\frac{1}{l}|F(T_m(\mathcal{N}_m))|^2,\frac{1}{l}|F(T_i(\mathcal{N}_i))|^2}(x,y)}{P_{\frac{1}{l}|F(T_i(\mathcal{N}_i))|^2}(y)} \\
&= \frac{\frac{(l-i)!}{(m-i-1)!(l-m)!}\left(1-\frac{f_c(x)}{f_c(y)}\right)^{m-i-1}\frac{P_c(x)}{f_c(y)}\left(\frac{f_c(x)}{f_c(y)}\right)^{l-m}}{P_{\frac{1}{l}|F(T_i(\mathcal{N}_i))|^2}(y)}, \ x \leq y.
\end{aligned} \tag{40}$$

Without loss of generality, at $l - 1$ available sub-channels, (38) can be rewritten as

$$\frac{1}{l-1}\left|F\left(T'_m\left(\mathcal{N}_m\right)\right)\right|^2, \ m = 0,\ldots,(l-1)-1, \tag{41}$$

and (39) holds, the conditional PDF is expressed as

$$P_{\frac{1}{l-1}|F(T'_m(\mathcal{N}_m))|^2}(x) = P_{\frac{1}{l}|F(T_m(\mathcal{N}_m))|^2 \Big| \frac{1}{l}|F(T_i(\mathcal{N}_i))|^2 = y}(x), \tag{42}$$

where the PDF of $\frac{1}{l-1}\left|F\left(T'_m\left(\mathcal{N}_m\right)\right)\right|^2$ is

$$P_{\frac{1}{l-1}|F(T'_m(\mathcal{N}_m))|^2}(x) = \frac{P_c(x)}{1-f_c(y)}, \ x \geq y. \tag{43}$$

In particular, focusing on the case that the $l$ sub-channels are selected via operator $\Lambda$ from a set of $n$ variables $S = \left\{\frac{1}{l}\left|F\left(T_i\left(\mathcal{N}_i\right)\right)\right|^2\right\}_{i=0}^{n-1}$, the joint PDF function is yielded as

$$\begin{aligned}
P_S\left(x_0,\ldots,x_{l-1}\right) &= \frac{n!}{(n-l)!}f_c\left(x_{l-1}\right)^{n-l}\prod_{i=0}^{l-1}P_c\left(x_i\right), \ x_0 \geq x_1 \geq \ldots \geq x_{l-1}.
\end{aligned} \tag{44}$$

The model of the $\Lambda_{U_k}$ order-and-sum (selection) operator of user $U_k$ with single-carrier channel $\mathcal{N}_{U_k}$ is depicted in Fig. 1. Operator $\Lambda_{U_k}$ is decomposed as $\Lambda_{U_k} = O\Sigma$, where $\Lambda_{U_k}$ first selects $l$ sub-channels from the total $n$ via an ordering operation $O$, which uses the normalized, unordered $\frac{1}{l}\left|F\left(\widehat{T}_i\left(\mathcal{N}_i\right)\right)\right|^2$ coefficients; then $\Lambda_{U_k}$ evaluates $A_j\left(\mathcal{N}_{U_k}\right) = (1/l)\sum_{i=0}^{l-1}F\left(T_i\left(\mathcal{N}_i\right)\right)$ from the ordered coefficients via the sum-operator $\Sigma$ to yield $A_j\left(\mathcal{N}_{U_k}\right)$.

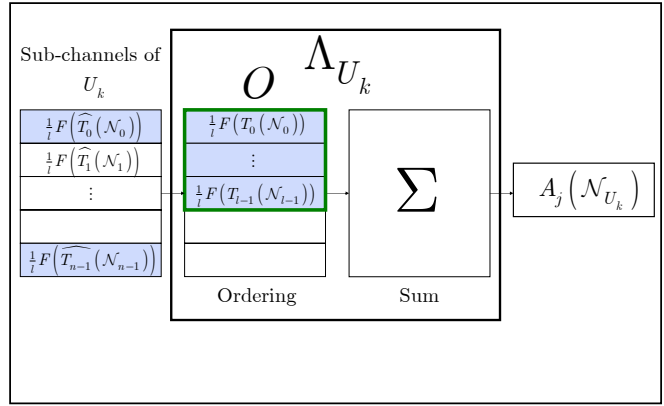

**Figure 1.** The $\Lambda_{U_k}$ order-and-sum operator of user $U_k$. $\Lambda_{U_k}$ selects $l$ sub-channels from the total $n$, via $O$. (The ordered $l$ sub-channels are depicted by the thick frame); then evaluates $A_j(\mathcal{N}_{U_k})$ of $\mathcal{N}_{U_k}$ via $\Sigma$.

### 3.2.2 Partial Sum Distributions of Ordered Sub-channels

**Proposition 1** (PDF of the partial sum of sub-channel coefficients). *The PDF function of single carrier transmittance* $\left|A_j(\mathcal{N}_j)\right|^2 = \tfrac{1}{l}\sum_{i=0}^{l-1}\left|F(T_i(\mathcal{N}_i))\right|^2$ *is* $P_{\left|A_j(\mathcal{N}_j)\right|^2}(x) = P_{\Gamma_1,\Gamma_2}(x-y,y)\,dy$, *where* $\Gamma_1 = \tfrac{1}{l}\sum_{i=0}^{l-2}\left|F(T_i(\mathcal{N}_i))\right|^2$ *and* $\Gamma_2 = \tfrac{1}{l}\left|F(T_{l-1}(\mathcal{N}_{l-1}))\right|^2$ *are two random correlated variables that identify the normalized, ordered sub-channel coefficients such that* $\tfrac{1}{l}\left|F(T_g(\mathcal{N}_i))\right|^2 \geq \tfrac{1}{l}\left|F(T_{g+1}(\mathcal{N}_{g+1}))\right|^2 \geq \tfrac{1}{l}\left|F(T_{l-1}(\mathcal{N}_{l-1}))\right|^2$, $g = 0,\ldots,l-2$, *where* $\left|F(T_{l-1}(\mathcal{N}_{l-1}))\right|^2$ *is the minimum transmittance of the l sub-channels.*

*Proof.*
Let $\left|A_j(\mathcal{N}_j)\right|^2$ be identified as
$$\left|A_j(\mathcal{N}_j)\right|^2 = \tfrac{1}{l}\sum_{i=0}^{l-1}\left|F(T_i(\mathcal{N}_i))\right|^2, \tag{45}$$
and let the transmittances to be sorted in a descending order as
$$\tfrac{1}{l}\left|F(T_0(\mathcal{N}_0))\right|^2 \geq \tfrac{1}{l}\left|F(T_1(\mathcal{N}_1))\right|^2 \ldots \geq \tfrac{1}{l}\left|F(T_{l-1}(\mathcal{N}_{l-1}))\right|^2. \tag{46}$$
Particularly, the random variable in (45) can be decomposed to the sum of two correlated random variables as
$$\left|A_j(\mathcal{N}_j)\right|^2 = \Gamma_1 + \Gamma_2, \tag{47}$$
where
$$\Gamma_1 = \tfrac{1}{l}\sum_{i=0}^{l-2}\left|F(T_i(\mathcal{N}_i))\right|^2 \tag{48}$$
and

$$\Gamma_2 = \tfrac{1}{l}\left|F\left(T_{l-1}\left(\mathcal{N}_{l-1}\right)\right)\right|^2. \tag{49}$$

Precisely, from (47) follows that the $P_{|A_j(\mathcal{N}_j)|^2}$ PDF function of $\left|A_j\left(\mathcal{N}_j\right)\right|^2$ can be evaluated via two random correlated variables as

$$P_{|A_j(\mathcal{N}_j)|^2}(x) = P_{\Gamma_1,\Gamma_2}(x-y, y)\, dy. \tag{50}$$

Specifically, exploiting the Bayesian formula, it follows that

$$P_{\Gamma_1,\Gamma_2}(x,y) = P_{\Gamma_2}(x) \times P_{\Gamma_1|\Gamma_2 = x}(y), \tag{51}$$

where $P_{\Gamma_2}(x)$ is the PDF of the $l$-th (ordered) sub-channel transmittance coefficient while $P_{\Gamma_1|\Gamma_2=x}(y)$ is the conditional PDF of the sum of the first $l$-1 ordered sub-channel coefficients, at $\tfrac{1}{l}\left|F\left(T_{l-1}\left(\mathcal{N}_{l-1}\right)\right)\right|^2 = x$.

Then let us assume that there are $l-1$ available sub-channels, and let the $i$-th sub-channel's quantity be referred to as $\tfrac{1}{l-1}\left|F\left(T_i'(\mathcal{N}_i)\right)\right|^2$, $i=0,\dots,l-2$.

Without loss of generality, for this quantity the following relation holds:

$$P_{\Gamma_1,\Gamma_2=y}(x) = P_{\tfrac{1}{l-1}\sum_{i=0}^{l-2}|F(T_i'(\mathcal{N}_i))|^2}(x), \tag{52}$$

where $P_{\tfrac{1}{l-1}|F(T_i'(\mathcal{N}_i))|^2}(x)$ is as

$$P_{\tfrac{1}{l-1}|F(T_i'(\mathcal{N}_i))|^2}(x) = \frac{P_c(x)}{1-f_c(x)}, x \geq y. \tag{53}$$

∎

### 3.2.3 Sub-channel Selection with a Complete Scan

**Theorem 1** (Sub-channel selection, $l$ from $n$). *The $\mathrm{M}$-function of $\left|F\left(T_i\left(\mathcal{N}_i\right)\right)\right|^2$ of the Gaussian sub-channel $\mathcal{N}_i$ selection operator $\Lambda_0$ is*

$$\mathrm{M}^{\Lambda_0}_{\tfrac{1}{l}|F(T_i(\mathcal{N}_i))|^2}(x) = f\!\left(\tfrac{1}{l}\left|F\left(T_i^*(\mathcal{N}_i)\right)\right|^2\right)$$
$$+\int_{\tfrac{1}{l}|F(T_i^*(\mathcal{N}_i))|^2}^{\infty} P\!\left(\tfrac{1}{l}\left|F\left(T_i(\mathcal{N}_i)\right)\right|^2\right) e^{x\tfrac{1}{l}|F(T_i(\mathcal{N}_i))|^2} d\tfrac{1}{l}\left|F\left(T_i(\mathcal{N}_i)\right)\right|^2,$$

*where $\tfrac{1}{l}\left|F\left(T_i^*(\mathcal{N}_i)\right)\right|^2 \leq \tfrac{1}{l}\left|F\left(T_i(\mathcal{N}_i)\right)\right|^2$ is an threshold for all $\mathcal{N}_i$, while $f(\cdot)$ and $P(\cdot)$ are the CDF and PDF functions of the ordered $\tfrac{1}{l}\left|F\left(T_i(\mathcal{N}_i)\right)\right|^2$ coefficients. For $l$ Gaussian sub-channels with $\tfrac{1}{l}\left|F\left(T_i(\mathcal{N}_i)\right)\right|^2 \leq \tfrac{1}{l}\left|F\left(T_i^*(\mathcal{N}_i)\right)\right|^2$, $i=0,\dots,l-1$, the $\mathrm{M}$-function of $\left|A_j\left(\mathcal{N}_j\right)\right|^2$ is*

$$\mathrm{M}^{\Lambda_0}_{|A_j(\mathcal{N}_j)|^2}(x) = \prod_l f\!\left(\tfrac{1}{l}\left|F\left(T_i^*(\mathcal{N}_i)\right)\right|^2\right)$$
$$+\int_{\tfrac{1}{l}|F(T_i^*(\mathcal{N}_i))|^2}^{\infty} P\!\left(\tfrac{1}{l}\left|F\left(T_i(\mathcal{N}_i)\right)\right|^2\right) e^{x\tfrac{1}{l}|F(T_i(\mathcal{N}_i))|^2} d\tfrac{1}{l}\left|F\left(T_i(\mathcal{N}_i)\right)\right|^2.$$

*Proof.*

First let us assume that the operation of the selection of the $l$ sub-channels with the best $\left|T_i\left(\mathcal{N}_i\right)\right|^2$ coefficient from the total $n$, is depicted by operator $\Lambda_0$.

Assuming that the $l$ sub-channels are selected via $\Lambda_0$ at a $\left|F\left(T_i^*\left(\mathcal{N}_i\right)\right)\right|^2$ threshold, the $P^{\Lambda_0}_{\left|A_j(\mathcal{N}_j)\right|^2}$ PDF function can be expressed via the

$$P^{\Lambda_0}_{\frac{1}{l}\sum_{i=0}^{l-2}\left|F(T_i(\mathcal{N}_i))\right|^2,\frac{1}{l}\left|F(T_{l-1}(\mathcal{N}_{l-1}))\right|^2}\left(x-y,y\right)dy = P_{\Gamma_1,\Gamma_2}\left(x-y,y\right)dy \tag{54}$$

joint PDF function of the sub-channel $\frac{1}{l}F\left(T_i\left(\mathcal{N}_i\right)\right)$-s, as

$$P^{\Lambda_0}_{\left|A_j(\mathcal{N}_j)\right|^2}(x) = \int_0^\infty P_{\Gamma_1,\Gamma_2}\left(x-y,y\right)dy. \tag{55}$$

The apriori fixed $\left|F\left(T_i^*\left(\mathcal{N}_i\right)\right)\right|^2$ threshold is determined as

$$\tfrac{1}{l}\left|F\left(T_i\left(\mathcal{N}_i\right)\right)\right|^2 \geq \tfrac{1}{l}\left|F\left(T_i^*\left(\mathcal{N}_i\right)\right)\right|^2,\ i = 0,\ldots,l-1, \tag{56}$$

for all $l$ Gaussian sub-channels that carry valuable information.

The $\left|F\left(T_i^*\left(\mathcal{N}_i\right)\right)\right|^2$ threshold is derived from the security threshold $\nu_{Eve}=1/\lambda$ of the optimal Gaussian attack (see [2]), where $\lambda$ is the Lagrange multiplier as

$$\lambda = \left|F\left(T_\mathcal{N}^*\right)\right|^2 = \tfrac{1}{l}\sum_{i=0}^{l-1}\left|F\left(T_i^*\left(\mathcal{N}_i\right)\right)\right|^2 = \tfrac{1}{l}\sum_{i=0}^{l-1}\left|\sum_{k=0}^{l-1}T_k^* e^{\frac{-i2\pi ik}{n}}\right|^2, \tag{57}$$

while $T_\mathcal{N}^*$ is the expected transmittance of the $l$ sub-channels under an optimal Gaussian attack. For a detailed description, see [2].

Without loss of generality, let us assume that the sub-channel transmittance coefficients are ordered in a descending order of (46).

Then if $\tfrac{1}{l}\left|F\left(T_i\left(\mathcal{N}_i\right)\right)\right|^2 < \tfrac{1}{l}\left|F\left(T_i^*\left(\mathcal{N}_i\right)\right)\right|^2$, let

$$P'^{\Lambda_0}_{\tfrac{1}{l}\left|F(T_i(\mathcal{N}_i))\right|^2}\left(\tfrac{1}{l}\left|F\left(T_i\left(\mathcal{N}_i\right)\right)\right|^2\right) = f^{\Lambda_0}_{\tfrac{1}{l}\left|F(T_i(\mathcal{N}_i))\right|^2}\left(\tfrac{1}{l}\left|F\left(T_i^*\left(\mathcal{N}_i\right)\right)\right|^2\right)\delta_c\left(\tfrac{1}{l}\left|F\left(T_i\left(\mathcal{N}_i\right)\right)\right|^2\right), \tag{58}$$

where $\delta_c(\cdot)$ is a random function, while for $\tfrac{1}{l}\left|F\left(T_i\left(\mathcal{N}_i\right)\right)\right|^2 \geq \tfrac{1}{l}\left|F\left(T_i^*\left(\mathcal{N}_i\right)\right)\right|^2$,

$$P'^{\Lambda_0}_{\tfrac{1}{l}\left|F(T_i(\mathcal{N}_i))\right|^2}\left(\tfrac{1}{l}\left|F\left(T_i\left(\mathcal{N}_i\right)\right)\right|^2\right) = P^{\Lambda_0}_{\tfrac{1}{l}\left|F(T_i(\mathcal{N}_i))\right|^2}\left(\tfrac{1}{l}\left|F\left(T_i\left(\mathcal{N}_i\right)\right)\right|^2\right). \tag{59}$$

Specifically, after some calculations, the M-function of $\tfrac{1}{l}\left|F\left(T_i\left(\mathcal{N}_i\right)\right)\right|^2$ of sub-channel $\mathcal{N}_i$ is yielded without loss of generality as

$$\begin{aligned}\mathrm{M}^{\Lambda_0}_{\tfrac{1}{l}\left|F(T_i(\mathcal{N}_i))\right|^2}(x) = {} & f^{\Lambda_0}_{\tfrac{1}{l}\left|F(T_i(\mathcal{N}_i))\right|^2}\left(\tfrac{1}{l}\left|F\left(T_i^*\left(\mathcal{N}_i\right)\right)\right|^2\right) \\ & + \int_{\tfrac{1}{l}\left|F(T_i^*(\mathcal{N}_i))\right|^2}^\infty P^{\Lambda_0}_{\tfrac{1}{l}\left|F(T_i(\mathcal{N}_i))\right|^2}\left(\tfrac{1}{l}\left|F\left(T_i\left(\mathcal{N}_i\right)\right)\right|^2\right)e^{x\tfrac{1}{l}\left|F(T_i(\mathcal{N}_i))\right|^2}d\tfrac{1}{l}\left|F\left(T_i\left(\mathcal{N}_i\right)\right)\right|^2.\end{aligned} \tag{60}$$

For $l$ Gaussian sub-channels with $\frac{1}{l}|F(T_i(\mathcal{N}_i))|^2 \geq \frac{1}{l}|F(T_i^*(\mathcal{N}_i))|^2, i = 0,\ldots,l-1$, the M-function of $|A_j(\mathcal{N}_j)|^2$ is

$$\begin{aligned}
\mathrm{M}^{\Lambda_0}_{|A_j(\mathcal{N}_j)|^2}(x) &= \prod_l \mathrm{M}^{\Lambda_0}_{\frac{1}{l}|F(T_i(\mathcal{N}_i))|^2}(x) \\
&= \prod_l f^{\Lambda_0}_{\frac{1}{l}|F(T_i(\mathcal{N}_i))|^2}\left(\frac{1}{l}|F(T_i^*(\mathcal{N}_i))|^2\right) \\
&+ \int_{\frac{1}{l}|F(T_i^*(\mathcal{N}_i))|^2}^{\infty} P^{\Lambda_0}_{\frac{1}{l}|F(T_i(\mathcal{N}_i))|^2}\left(\frac{1}{l}|F(T_i(\mathcal{N}_i))|^2\right) e^{x\frac{1}{l}|F(T_i(\mathcal{N}_i))|^2} d\frac{1}{l}|F(T_i(\mathcal{N}_i))|^2.
\end{aligned} \tag{61}$$

In particular, the $\mathrm{Pr}_{\Lambda_0}(\cdot)$ probability that $l$ sub-channels are selected from the $n$ is as

$$\mathrm{Pr}_{\Lambda_0}(l) = \binom{n}{l}\left(1 - f_c\left(\frac{1}{l}|F(T_i^*(\mathcal{N}_i))|^2\right)\right)^l \left(f_c\left(\frac{1}{l}|F(T_i^*(\mathcal{N}_i))|^2\right)\right)^{n-l}. \tag{62}$$

Note that in a worst-case scenario, the condition $\frac{1}{l}|F(T_i(\mathcal{N}_i))|^2 \geq \frac{1}{l}|F(T_i^*(\mathcal{N}_i))|^2$ may not be satisfied for the required number $l$ of sub-channels. In this case, the threshold $\frac{1}{l}|F(T_i^*(\mathcal{N}_i))|^2$ can be redefined as

$$\frac{1}{l}|F(T_i^*(\mathcal{N}_i))|^2 = \mu \max_l \frac{1}{l}|(F(T_i(\mathcal{N}_i)))|^2, \tag{63}$$

where $0 < \mu < 1$ is a real variable. The condition of (63) satisfies that at least, the best sub-channel is selected via the sub-channel allocation procedure. It also allows us to reevaluate the M-function of the $l$ sub-channels as follows.

Presuming that for $l$ sub-channels, (63) is satisfied, resulting the ordered coefficients $\frac{1}{l}|F(T_i(\mathcal{N}_i))|^2 \geq \frac{1}{l}|F(T_{i+1}(\mathcal{N}_{i+1}))|^2$. Using the $P^{\Lambda_0}_{\frac{1}{l}|F(T_0(\mathcal{N}_0))|^2 \ldots \frac{1}{l}|F(T_{l-1}(\mathcal{N}_{l-1}))|^2}(\cdot)$ joint PDF function, the $\mathrm{M}^{\Lambda_0}_{|A_j(\mathcal{N}_j)|^2}(x)$-function is precisely as

$$\begin{aligned}
\mathrm{M}^{\Lambda_0}_{|A_j(\mathcal{N}_j)|^2}(x) &= \int_0^{\infty} d_{\frac{1}{l}|F(T_0(\mathcal{N}_0))|^2} \int_{\mu\frac{1}{l}|F(T_0(\mathcal{N}_0))|^2}^{\frac{1}{l}|F(T_0(\mathcal{N}_0))|^2} d_{\frac{1}{l}|F(T_1(\mathcal{N}_1))|^2} \\
&\cdots \int_{\mu\frac{1}{l}|F(T_0(\mathcal{N}_0))|^2}^{\frac{1}{l}|F(T_{l-2}(\mathcal{N}_{l-2}))|^2} d_{\frac{1}{l}|F(T_{l-1}(\mathcal{N}_{l-1}))|^2} \int_0^{\mu\frac{1}{l}|F(T_0(\mathcal{N}_0))|^2} d_{\frac{1}{l}|F(T_l(\mathcal{N}_l))|^2} \\
&\int_0^{\frac{1}{l}|F(T_l(\mathcal{N}_l))|^2} d_{\frac{1}{l}|F(T_{l+1}(\mathcal{N}_{l+1}))|^2} \\
&\cdots \int_0^{\mu\frac{1}{l}|F(T_{n-1}(\mathcal{N}_{n-1}))|^2} e^{x\frac{1}{l}\sum_{i=0}^{l-1}|F(T_i(\mathcal{N}_i))|^2} \\
&\times P\left(\frac{1}{l}|F(T_0(\mathcal{N}_0))|^2 \ldots \frac{1}{l}|F(T_{n-1}(\mathcal{N}_{n-1}))|^2\right) d\frac{1}{l}|F(T_{n-1}(\mathcal{N}_{n-1}))|^2,
\end{aligned} \tag{64}$$

which can be simplified into

$$\mathrm{M}^{\Lambda_0}_{|A_j(\mathcal{N}_j)|^2}(x) = l\binom{n}{l}\int_0^\infty e^{x\frac{1}{l}|F(T_i(\mathcal{N}_i))|^2} P\left(\frac{1}{l}|F(T_i(\mathcal{N}_i))|^2\right)$$
$$\cdot \left(f\left(\mu\frac{1}{l}|F(T_i(\mathcal{N}_i))|^2\right)\right)^{n-l} \left(\mathrm{M}\left(x,\frac{1}{l}|F(T_i(\mathcal{N}_i))|^2\right)\right.$$
$$\left. - \mathrm{M}\left(x,\mu\frac{1}{l}|F(T_i(\mathcal{N}_i))|^2\right)\right)^{l-1} d\frac{1}{l}|F(T_i(\mathcal{N}_i))|^2, \tag{65}$$

and the partial function is
$$\mathrm{M}^{\Lambda_0}_{\frac{1}{l}|F(T_i(\mathcal{N}_i))|^2}(w,x)$$
$$= \int_x^\infty P\left(\frac{1}{l}|F(T_i(\mathcal{N}_i))|^2\right) e^{w\frac{1}{l}|F(T_i(\mathcal{N}_i))|^2} d\frac{1}{l}|F(T_i(\mathcal{N}_i))|^2, \tag{66}$$

by some fundamental theory.

The probability that $k$ sub-channels are selected from $n$ at condition (63), is expressed as
$$\Pr\nolimits_{\Lambda_0}(k) = \Pr\left(\frac{1}{l}|F(T_k(\mathcal{N}_k))|^2 \geq \frac{1}{l}|F(T_i^*(\mathcal{N}_i))|^2 \geq \frac{1}{l}|F(T_{k+1}(\mathcal{N}_{k+1}))|^2\right)$$
$$= \int_0^\infty \int_{\mu x}^x \int_0^{\mu x} P_{0,k,k+1}(x,y,z)\, dx dy dz, \tag{67}$$

where
$$P_{0,k-1,k}(x,y,z) = \frac{n!}{(k-2)!(n-k-1)!} P_c(x)$$
$$\cdot (f_c(x) - f_c(y))^{k-2} P_c(y) \tag{68}$$
$$\cdot P_c(z)(f_c(z))^{n-k-1}.$$

■

## 3.3 Optimized Complexity Progressive Scan

**Theorem 2** (Progressive sub-channel selection). *The $\kappa$ average number of iterations needed for the selection of the $l$ Gaussian sub-channels is minimized via an $\Lambda$ sub-channel selection operator.*

*Proof.*
The complexity of the sub-channel selection operators is discusses via the $\kappa$ average number of iterations needed for the procedure. The proof follows the definitions and notations of [29].
Exploiting the results of Theorem 1, using $\Lambda_0$, the $\kappa_{\Lambda_0}$ overall average number of the iterations (e.g., the number of comparisons of sub-channel transmittance coefficients at a $|F(T_i^*(\mathcal{N}_i))|$ threshold per sub-channels) needed to derive $|A_j|^2$ is yielded as
$$\kappa_{\Lambda_0} = \sum_{i=1}^l \binom{n}{i} \left(1 - f_c\left(\frac{1}{l}|F(T_i^*(\mathcal{N}_i))|^2\right)\right)^i \left(f_c\left(\frac{1}{l}|F(T_i^*(\mathcal{N}_i))|^2\right)\right)^{n-i}$$
$$\cdot \left(n + \sum_{k=1}^{l-i}(n-i-k)\right). \tag{69}$$

Operator $\Lambda_0$ does require the scan of the total $n$ sub-channels, which is practically inconvenient. To resolve this problem we introduce operator $\Lambda$ which performs a *progressive* scan: it stops the iteration as the $l$ sub-channels have found and does not require to scan through all the $n$ sub-channels. Let $\Lambda'$ a slightly modified version of $\Lambda$ that also handles the situation when only $k < l$ sub-channels are found; but $l$ was required by the legal parties for the transmission. In this case, the progressive scan operator $\Lambda'$ selects the remaining $l-k$ sub-channels from the set $\mathcal{B}(\mathcal{N}_k,\ldots,\mathcal{N}_{l-1})$ of bad sub-channels, and the Lagrange multiplier in (57) at $\Lambda'$ is reevaluated as

$$\lambda_{\Lambda'} = \left|F\left(\widehat{T}_\mathcal{N}\right)\right|^2 = \left|F\left(T_i^*(\mathcal{N}_i)\right)\right|^2 + \varpi, \qquad (70)$$

where $\varpi$ is a nonnegative real variable, expressed as

$$\varpi = \left|F\left(T_i^*(\mathcal{N}_i)\right)\right|^2 - \min\left(\left|F\left(T(\mathcal{N}_k)\right)\right|^2,\ldots,\left|F\left(T(\mathcal{N}_{l-1})\right)\right|^2\right), \qquad (71).$$

where $\left|F\left(T(\mathcal{N}_k)\right)\right|^2,\ldots,\left|F\left(T(\mathcal{N}_{l-1})\right)\right|^2$ are the corresponding coefficients of $\mathcal{B}(\mathcal{N}_k,\ldots,\mathcal{N}_{l-1})$. As follows, (70) allows to the legal parties to preserve the security conditions via a modified security threshold $\nu'_{Eve} = 1/\lambda_{\Lambda'}$.

The $\kappa_\Lambda$ and $\kappa_{\Lambda'}$ average number of iterations of operators $\Lambda$ and $\Lambda'$ are evaluated as follows. Without loss generality, the $\Lambda$ sub-channel selection operator models the situation if the iteration stops as the $\tfrac{1}{l}\left|F\left(T_i(\mathcal{N}_i)\right)\right|^2 \geq \tfrac{1}{l}\left|F\left(T_i^*(\mathcal{N}_i)\right)\right|^2$ condition is satisfied for $l$ sub-channels, that is, it is not needed to determine $\tfrac{1}{l}\left|F\left(T_i(\mathcal{N}_i)\right)\right|^2$ for the remaining $i = l,\ldots,n-1$ sub-channels. Specifically, the $\Pr_\Lambda(k)$ probability of the availability of $l$ sub-channels for $k = l$,

$$\Pr_\Lambda(k) = \sum_{h=l}^{n} \binom{n}{h}\left(1 - f_c\left(\tfrac{1}{l}\left|F\left(T_h^*(\mathcal{N}_h)\right)\right|^2\right)\right)^h \cdot \left(f_c\left(\tfrac{1}{l}\left|F\left(T_h^*(\mathcal{N}_h)\right)\right|^2\right)\right)^{n-h}, \qquad (72)$$

and the probability that only $k < l$, $k = 0,\ldots,l-1$ sub-channels are available is

$$\Pr_\Lambda(k) = \binom{n}{k}\left(1 - f_c\left(\tfrac{1}{l}\left|F\left(T_i^*(\mathcal{N}_i)\right)\right|^2\right)\right)^k \cdot \left(f_c\left(\tfrac{1}{l}\left|F\left(T_i^*(\mathcal{N}_i)\right)\right|^2\right)\right)^{n-k}, \qquad (73)$$

and

$$\mathrm{M}^\Lambda_{|A_j(\mathcal{N}_j)|^2}(x) = \sum_{h=0}^{l} \Pr_\Lambda(h)\mathrm{M}^{\Lambda(h)}_{|A_j(\mathcal{N}_j)|^2}(x), \qquad (74)$$

where $\mathrm{M}^{\Lambda(h)}_{|A_j(\mathcal{N}_j)|^2}(x)$ is the conditional MGF at $i$ available Gaussian sub-channels.

Particularly, (74) can be rewritten precisely as

$$M^{\Lambda}_{|A_j(\mathcal{N}_j)|^2}(x) = \sum_{h=0}^{l-1} \binom{n}{h}\left(1-P\left(\tfrac{1}{l}\left|F\left(T_h^*(\mathcal{N}_h)\right)\right|^2\right)\right)^h$$
$$\cdot \left(P\left(\tfrac{1}{l}\left|F\left(T_h^*(\mathcal{N}_h)\right)\right|^2\right)\right)^{n-h} M^{\Lambda(h)}_{|A_j(\mathcal{N}_j)|^2}(x)$$
$$+ \sum_{g=l}^{n} \binom{n}{g}\left(1-P\left(\tfrac{1}{l}\left|F\left(T_g^*(\mathcal{N}_i)\right)\right|^2\right)\right)^g \quad (75)$$
$$\cdot \left(P\left(\tfrac{1}{l}\left|F\left(T_g^*(\mathcal{N}_i)\right)\right|^2\right)\right)^{n-g} \frac{M^{\Lambda(h)}_{|A_j(\mathcal{N}_j)|^2}(x)}{\left(M^{\Lambda(2)}_{|A_j(\mathcal{N}_j)|^2}\right)^{l-i}},$$

where

$$M^{\Lambda(h)}_{|A_j(\mathcal{N}_j)|^2}(x) = \left(M^{\Lambda(1)}_{|A_j(\mathcal{N}_j)|^2}(x)\right)^h \left(M^{\Lambda(2)}_{|A_j(\mathcal{N}_j)|^2}(x)\right)^{l-h} \quad (76)$$

which can be rewritten as

$$M^{\Lambda}_{|A_j(\mathcal{N}_j)|^2}(x) = \left(1 - f_c\left(\tfrac{1}{l}\left|F\left(T_i^*(\mathcal{N}_i)\right)\right|^2\right)\right) M^{\Lambda(1)}_{|A_j(\mathcal{N}_j)|^2}(x)$$
$$+ f_c\left(\tfrac{1}{l}\left|F\left(T_i^*(\mathcal{N}_i)\right)\right|^2\right) M^{\Lambda(2)}_{|A_j(\mathcal{N}_j)|^2}(x). \quad (77)$$

After some calculations, the $\kappa_\Lambda$ overall average number of the iterations to determine $A_j$ is as

$$\kappa_\Lambda = \sum_{i=l}^{n} i\left(1 - f_c\left(\tfrac{1}{l}\left|F\left(T_i^*(\mathcal{N}_i)\right)\right|^2\right)\right)^l \binom{i-1}{i-l}$$
$$\cdot \left(f_c\left(\tfrac{1}{l}\left|F\left(T_i^*(\mathcal{N}_i)\right)\right|^2\right)\right)^{i-l}$$
$$+ n \sum_{n-l+1}^{n} \binom{n}{i}\left(1 - f_c\left(\tfrac{1}{l}\left|F\left(T_i^*(\mathcal{N}_i)\right)\right|^2\right)\right)^i \quad (78)$$
$$\cdot \left(f_c\left(\tfrac{1}{l}\left|F\left(T_i^*(\mathcal{N}_i)\right)\right|^2\right)\right)^{n-i}.$$

Assuming the situation that the $k$ available number of sub-channels is lower than the expected $l$, the best $l-k$ "bad" sub-channels with $\tfrac{1}{l}\left|F\left(T_i(\mathcal{N}_i)\right)\right|^2 < \tfrac{1}{l}\left|F\left(T_i^*(\mathcal{N}_i)\right)\right|^2, i = k,\dots,l-1$ can be selected for transmission. This change is modeled via an extended $\Lambda'$ operator that does modify the complexity and also leads to different functions and operators. Without loss of generality, if only $k < l$ "good" sub-channels, $\mathcal{G} = \{\mathcal{N}_0,\dots,\mathcal{N}_{k-1}\}$, are found and the remaining $l-k$ sub-channels are selected set $\mathcal{B}$ of the ordered coefficients, the $f^{\Lambda'}_{|A_j(\mathcal{N}_j)|^2}$ CDF-function is as

$$f^{\Lambda'}_{|A_j(\mathcal{N}_j)|^2}(x) = \Pr\left(\left|A_j(\mathcal{N}_j)\right|^2 < x\right)$$
$$= \Pr\nolimits_{\Lambda'}(l)\Pr\nolimits_{\Lambda'}\left(\left|A_j(\mathcal{N}_j)\right|^2 < x \big| k \geq l\right) \quad (79)$$
$$+ \Pr\nolimits_{\Lambda'}\left(\left|A_j(\mathcal{N}_j)\right|^2 < x, k < l\right),$$

where $\Pr_{\Lambda'}(l) = \Pr_{\Lambda}(l) = \Pr_{\Lambda_0}(l)$.

Since at $k \geq l$, $|A_j(\mathcal{N}_j)|^2 > \frac{1}{l}\sum_{i=0}^{l-1}|F(T_i^*(\mathcal{N}_i))|^2 = |F(T_i^*(\mathcal{N}_i))|^2$, it follows that

$$\Pr_{\Lambda'}\left(|A_j(\mathcal{N}_j)|^2 < x \big| k \geq l\right)$$
$$= 1 - e^{-\frac{x-|F(T_i^*(\mathcal{N}_i))|^2}{|A_j(\mathcal{N}_j)|^2}} \sum_{t=0}^{l-1} \frac{1}{t!}\left(\frac{x-|F(T_i^*(\mathcal{N}_i))|^2}{|A_j(\mathcal{N}_j)|^2}\right)^k, x \geq |F(T_i^*(\mathcal{N}_i))|^2, \quad (80)$$

while for $k < l$ the corresponding relation for $\Pr_{\Lambda'}\left(|A_j(\mathcal{N}_j)|^2 < x, k < l\right)$ is

$$\Pr_{\Lambda'}\left(|A_j(\mathcal{N}_j)|^2 < x, k < l\right)$$
$$= \Pr_{\Lambda'}\left(\Gamma_1 + \Gamma_2 < x, \Gamma_2 < \tfrac{1}{l}|F(T_i^*(\mathcal{N}_i))|^2\right), \quad (81)$$

where $\Gamma_1, \Gamma_2$ were defined in (48) and (49), respectively, and

$$\Pr_{\Lambda'}\left(\Gamma_1 + \Gamma_2 < x, \Gamma_2 < \tfrac{1}{l}|F(T_i^*(\mathcal{N}_i))|^2\right)$$
$$= \int_0^{\min\left(\tfrac{1}{l}|F(T_i^*(\mathcal{N}_i))|^2, \tfrac{x}{l}\right)} \int_{(l-2)y}^{x-y} P_{\Gamma_2, \Gamma_1}(y, z)\, dz\, dy. \quad (82)$$

After some calculations, the $\kappa_{\Lambda'}$ overall average number of the iterations at $\Lambda'$ is yielded as

$$\kappa_{\Lambda'} = \sum_{i=l}^{n} i \binom{i-1}{i-l}\left(1 - f_c\left(\tfrac{1}{l}|F(T_i^*(\mathcal{N}_i))|^2\right)\right)^l$$
$$\cdot \left(f_c\left(\tfrac{1}{l}|F(T_i^*(\mathcal{N}_i))|^2\right)\right)^{i-l}$$
$$+ \sum_{k=0}^{l-1}\left(n + \sum_{i=1}^{l-k}(n-k-i)\right)\binom{n}{k} f_c\left(\tfrac{1}{l}|F(T_i^*(\mathcal{N}_i))|^2\right)^{n-k} \quad (83)$$
$$\cdot \left(1 - f_c\left(\tfrac{1}{l}|F(T_i^*(\mathcal{N}_i))|^2\right)\right)^k.$$

The values of parameter $\kappa$ from (69), (78) and (83) are compared in Fig. 2.

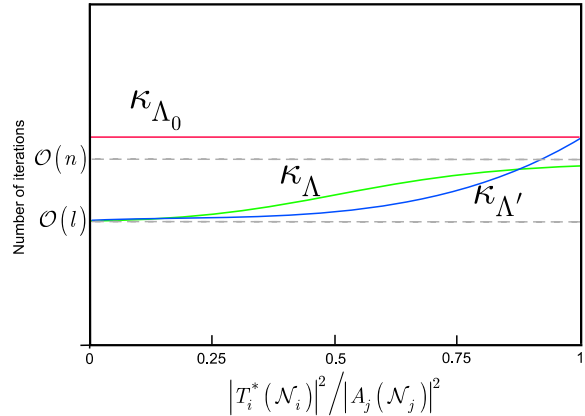

**Figure 2.** Comparison of $\kappa_{\Lambda_0}$, $\kappa_{\Lambda}$, and $\kappa_{\Lambda'}$ in function of $|T_i^*(\mathcal{N}_i)|^2 / |A_j(\mathcal{N}_j)|^2$.

The number of iterations at $\Lambda_0$ is $\kappa_{\Lambda_0} > \mathcal{O}(n)$ for all values of $\left|T_i^*(\mathcal{N}_i)\right|^2 \Big/ \left|A_j(\mathcal{N}_j)\right|^2$. Increasing $\left|T_i^*(\mathcal{N}_i)\right|^2 \Big/ \left|A_j(\mathcal{N}_j)\right|^2$, $\kappa_\Lambda$ converges from $\mathcal{O}(l)$ to $\mathcal{O}(n)$, while $\kappa_{\Lambda'}$ increases from $\mathcal{O}(l)$ to $\kappa_{\Lambda_0}$. The reason is that at $\Lambda_0$ all $n$ sub-channels are scanned (non-progressive operator). Using $\Lambda$, the progressive operator selects the first $l$ sub-channels that requires a lower number of iterations. At $\Lambda'$, for low $\left|T_i^*(\mathcal{N}_i)\right|^2 \Big/ \left|A_j(\mathcal{N}_j)\right|^2$ the first $l$ sub-channels were selected, which results in $\kappa_{\Lambda'} \approx \kappa_\Lambda \approx \mathcal{O}(l)$, while as $\left|T_i^*(\mathcal{N}_i)\right|^2 \Big/ \left|A_j(\mathcal{N}_j)\right|^2$ increases, the iteration is extended to all $n$ sub-channels that yields $\kappa_{\Lambda'} \approx \mathcal{O}(n)$.

∎

## 3.4 Distribution of Gaussian Subcarriers

**Proposition 2** (Distribution of the Gaussian subcarriers). *The (normalized) $\partial_i = \frac{1}{l}\left|d_i'\right|^2$ squared subcarrier magnitude is exponentially distributed, with single-carrier squared magnitude $\alpha_j(\mathcal{N}_j) = \sum_{i=0}^{l-1} \partial_i = \frac{1}{l}\sum_{i=0}^{l-1}\left|d_i'\right|^2$, where $d_i' \in \mathcal{CN}(0, \sigma_{\omega_0}^2)$ is the noisy Gaussian subcarrier.*

*Proof.*
Let the output of $\mathcal{N}_i$ be the noisy Gaussian subcarrier $d_i' \in \mathcal{CN}(0, \sigma_{\omega_0}^2)$. Then the normalized random variable

$$\partial_i = \tfrac{1}{l}\left|d_i'\right|^2 \tag{84}$$

is exponentially distributed with density

$$f(\partial_i) = \tfrac{1}{\sigma_{\partial_i}^2} e^{\frac{-\partial_i}{\sigma_{\partial_i}^2}}, \ \partial_i \geq 0, \tag{85}$$

which can be rewritten as

$$f(x) = \tfrac{1}{\tilde{\partial}} e^{\frac{-x}{\tilde{\partial}}}, \ x \geq 0, \tag{86}$$

where $\tilde{\partial}$ is the common mean.
The PDF of $P_\tau(x) = P_{\partial_i}(x) \big/ f_{\partial_i}(\partial_i), 0 < x < \partial_i$, is evaluated as

$$P_\tau(x) = \frac{P_{\partial_i}(x)}{f_{\partial_i}(\partial_i)} = \frac{\tfrac{1}{\tilde{\partial}} e^{\frac{-x}{\tilde{\partial}}}}{1 - e^{\frac{-\partial_i}{\tilde{\partial}}}}, \tag{87}$$

while $P_\zeta(x) = P_{\partial_i}(x) \big/ \left(f_{\partial_i}(\partial_i) - f_{\partial_i}(\delta)\right), \delta \leq x \leq \partial_i$, is expressed as

$$P_\zeta(x) = \frac{P_{\partial_i}(x)}{f_{\partial_i}(\partial_i) - f_{\partial_i}(\delta)} = \frac{e^{\frac{-x}{\tilde{\partial}}}}{\tilde{\partial}\left(e^{\frac{-\delta}{\tilde{\partial}}} - e^{\frac{-\partial_i}{\tilde{\partial}}}\right)}, \tag{88}$$

with an M-function

$$\begin{aligned}\mathrm{M}_\zeta(w) &= \int_\delta^{\partial_i} P_\zeta(x)e^{wx}dx \\ &= \frac{e^{w\delta-\delta/\tilde{\partial}}-e^{w\partial_i-\partial_i/\tilde{\partial}}}{\left(e^{-\delta/\tilde{\partial}}-e^{-\partial_i/\tilde{\partial}}\right)\left(1-w\tilde{\partial}\right)}.\end{aligned} \quad (89)$$

Then let $\alpha_j(\mathcal{N}_j)$ be the averaged single-carrier squared magnitude as the sum of $\partial_i$-s, as

$$\begin{aligned}\alpha_j(\mathcal{N}_j) &= \sum_{i=0}^{l-1}\partial_i \\ &= \frac{1}{l}\sum_{i=0}^{l-1}\left|d_i'\right|^2,\end{aligned} \quad (90)$$

which can be rewritten as the sum of independent random variables as

$$\alpha_j(\mathcal{N}_j) = \sum_{i=1}^{l}\sum_{k=i}^{n} x_k = x_1 + 2x_2 + \ldots + lx_l + lx_{l+1} + \ldots lx_n. \quad (91)$$

Specifically, using (91), the M-function of a term $kx_k$ is yielded as

$$\begin{aligned}\mathrm{M}_{kx_h}(w) &= \int_0^\infty P_{kx_h}(x)e^{wx}dx \\ &= \left(1-\frac{w\tilde{\partial}k}{h}\right)^{-1},\end{aligned} \quad (92)$$

and the M-function of $\alpha_j(\mathcal{N}_j)$ is as

$$\mathrm{M}_{\alpha_j(\mathcal{N}_j)}(w) = \left(1-w\tilde{\partial}\right)^{-l}\prod_{k=l+1}^{n}\left(1-\frac{w\tilde{\partial}l}{k}\right)^{-1}. \quad (93)$$

Then $P_{\alpha_j(\mathcal{N}_j)}(x)$ is expressed as

$$\begin{aligned}P_{\alpha_j(\mathcal{N}_j)}(x) &= \frac{n!}{(n-l)!l!}e^{\frac{-x}{\tilde{\partial}}}\left[\frac{x^{l-1}}{\tilde{\partial}_i^l(l-1)!}+\frac{1}{\tilde{\partial}}\sum_{k=1}^{n-l}(-1)^{l+k-1}\right. \\ &\left.\cdot\frac{(n-l)!}{(n-l-k)!k!}\left(\frac{l}{k}\right)^{l-1}\cdot\left(e^{\frac{-kx}{l\tilde{\partial}}}-\sum_{m=0}^{l-2}\frac{1}{m!}\left(\frac{-kx}{l\tilde{\partial}}\right)^m\right)\right].\end{aligned} \quad (94)$$

Particularly, at $l-1$ total sub-channels, let us denote $\partial_i' = \frac{1}{l-1}\left|d_i'\right|^2$ the $i$-th normalized Gaussian subcarrier. Then

$$P_{\partial_i'}(y) = \frac{1}{\tilde{\partial}}e^{\frac{-x-y}{\tilde{\partial}}}, \quad x \geq y, \quad (95)$$

and the conditional PDF is as

$$\begin{aligned}P_{\Gamma_1|\Gamma_2=y}(x) &= P_{\sum_{i=0}^{l-2}\partial_i'}(x) \\ &= \frac{1}{(l-2)!\tilde{\partial}^{l-1}}\left(x-(l-1)y\right)^{(l-2)}e^{-\frac{x-(l-1)y}{\tilde{\partial}}}, \quad x \geq (l-1)y,\end{aligned} \quad (96)$$

where $\Gamma_1 = \sum_{i=0}^{l-2}\partial_i$, $\Gamma_2 = \partial_{l-1}$, i.e., $\partial_i$-s are ordered such that $\partial_0 \geq \ldots \geq \partial_{l-2} \geq \partial_{l-1}$, while

$$M_{\partial'_i}(w) = \int_0^{+\infty} P_{\partial'_i}(y) e^{wx} dx \tag{97}$$
$$= \frac{1}{1-w\tilde{\partial}} e^{wy},$$

and the M-function of $\sum_{i=0}^{l-2} \partial'_i$ is

$$M_{\sum_{i=0}^{l-2} \partial'_i}(w) = \left(M_{\partial'_i}\right)^{l-1} \tag{98}$$
$$= \frac{1}{\left(1-w\tilde{\partial}\right)^{l-1}} e^{(l-1)wx}.$$

The joint PDF is

$$P_{\Gamma_2,\Gamma_1}(x,y) = \frac{n!}{(n-l)!(l-1)!} \left(f_c(x)\right)^{n-l} \tag{99}$$
$$\cdot \left(1 - f_c(x)\right)^{l-1} P_c(x) P_{\sum_{i=0}^{l-2} \partial'_i}(y),$$

where $P_c(x)$ and $f_c(x)$ are the common PDF and CDF of the unordered variables, while

$$P_{\Gamma_2}(x) = \frac{n!}{\tilde{\partial}(l-1)!} \sum_{h=0}^{n-l} \frac{(-1)^h}{(n-l-h)!h!} e^{-\frac{(l+h)x}{\tilde{\partial}}}. \tag{100}$$

Using (100), the joint PDF in (99) can be rewritten precisely as

$$P_{\Gamma_2,\Gamma_1}(x,y)$$
$$= \sum_{h=0}^{n-l} \frac{(-1)^h n!}{(n-l-h)!(l-1)!(l-2)!h!\tilde{\partial}^l} \left(y-(l-1)x\right)^{(l-2)} e^{\frac{-y+(h+1)x}{\tilde{\partial}}}, \; x \geq 0, y \geq (l-1)x. \tag{101}$$

Note that by some fundamental theory

$$\lim_{x \to +\infty} \frac{1-f_c(x)}{P_c(x)} = \Omega > 0, \tag{102}$$

and

$$\lim_{l \to +\infty} f_{\partial_0}(x) = e^{-e^{-x-\log l}}. \tag{103}$$

Then, the $x$ and $p$ quadrature-level (single-carrier) error probability at $\Lambda_0$ is as

$$p_{err}^{\Lambda_0}(A_j) = \frac{1}{\pi} \int_0^{\frac{\pi}{2}} M_{|A_j(\mathcal{N}_j)|^2 \cdot \widehat{SNR}}^{\Lambda_0} \left(\frac{\sin^2\left(\frac{\pi}{2}\right)}{\sin^2 \phi}\right) d\phi, \tag{104}$$

where $M_{|A_j(\mathcal{N}_j)|^2 \cdot \widehat{SNR}}^{\Lambda_0}$ is the M-function of $|A_j(\mathcal{N}_j)|^2 \cdot \widehat{SNR}$; $\widehat{SNR}$ is a scaled SNR quantity, $\widehat{SNR} = \sigma_{\omega_0}^2 / 2\sigma_{\mathcal{N}}^2$, and $\phi \in [0, 2\pi]$. In particular, the formula of (104) can be applied to derive the $p_{err}(A_j)$ error probabilities (single-carrier quadrature level) of the operators using the M-functions proposed in sub-sections 3.1-3.2, precisely as

$$p_{err}^{\Lambda}(A_j) = \frac{1}{\pi} \int_0^{\frac{\pi}{2}} M_{|A_j(\mathcal{N}_j)|^2 \cdot \widehat{SNR}}^{\Lambda} \left(\frac{\sin^2\left(\frac{\pi}{2}\right)}{\sin^2 \phi}\right) d\phi, \tag{105}$$

and

$$p_{err}^{\Lambda'}(A_j) = \frac{1}{\pi} \int_0^{\frac{\pi}{2}} M_{|A_j(\mathcal{N}_j)|^2 \cdot \widehat{SNR}}^{\Lambda'} \left(\frac{\sin^2\left(\frac{\pi}{2}\right)}{\sin^2 \phi}\right) d\phi. \tag{106}$$

Specifically, assuming $T_i(\mathcal{N}_i) \in \mathcal{CN}(0, \sigma^2_{T_i(\mathcal{N}_i)})$, and by using the M-functions derived in the previous sections, the corresponding $p_{err}^{\Lambda_0}(A_j)$ single-carrier quadrature level error probability at $\Lambda_0$ in function of $|A_j(\mathcal{N}_j)|^2 \cdot \widehat{\text{SNR}}$, are yielded as shown in Fig. 3. The results can be extended for arbitrarily distributed $T_i(\mathcal{N}_i)$ coefficients.

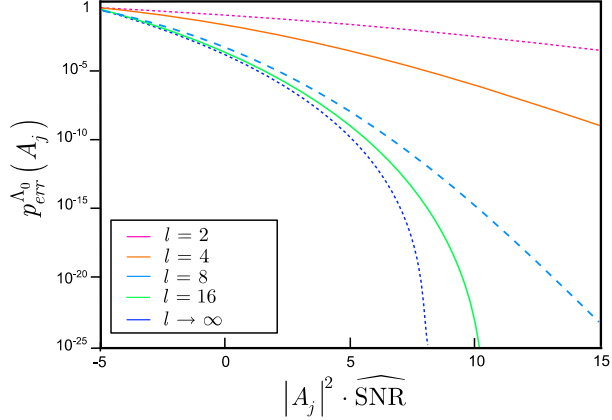

**Figure 3.** The $p_{err}^{\Lambda_0}(A_j)$ at $\Lambda_0$ via $\text{M}^{\Lambda_0}_{|A_j(\mathcal{N}_j)|^2 \cdot \widehat{\text{SNR}}}$, in function of $|A_j(\mathcal{N}_j)|^2 \cdot \widehat{\text{SNR}}$, $T_i(\mathcal{N}_i) \in \mathcal{CN}(0, \sigma^2_{T_i(\mathcal{N}_i)})$.

Using operators $\Lambda$ and $\Lambda'$ with $\text{M}^{\Lambda}_{|A_j(\mathcal{N}_j)|^2 \cdot \widehat{\text{SNR}}}(\cdot)$ and $\text{M}^{\Lambda'}_{|A_j(\mathcal{N}_j)|^2 \cdot \widehat{\text{SNR}}}$, the complexity can be reduced, however, the yielding single-carrier quadrature level error probabilities $p_{err}^{\Lambda}(A_j)$ and $p_{err}^{\Lambda'}(A_j)$ are slightly increased, as depicted in Fig. 4.

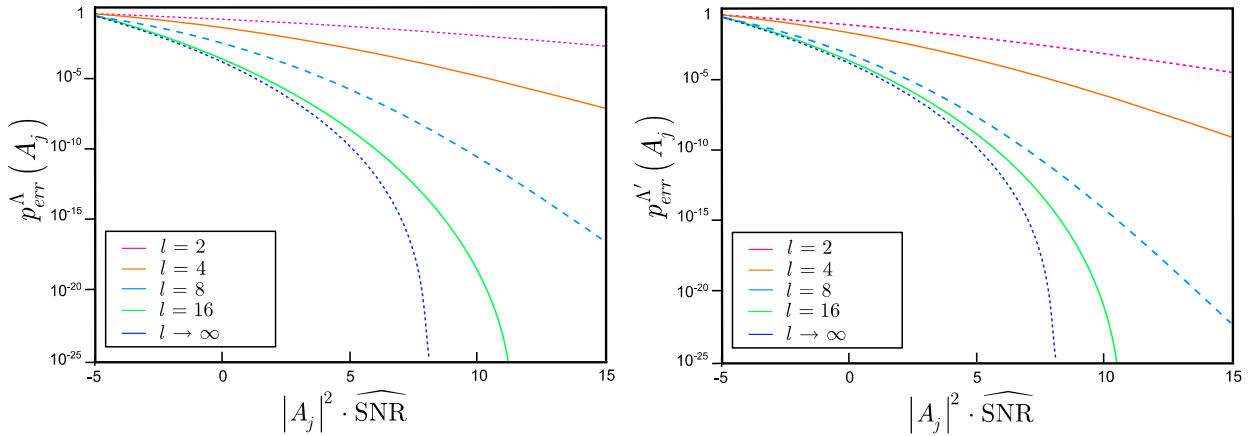

**Figure 4.** The $p_{err}^{\Lambda}(A_j)$ and $p_{err}^{\Lambda'}(A_j)$ at $\Lambda$ and $\Lambda'$ via $\text{M}^{\Lambda}_{|A_j(\mathcal{N}_j)|^2 \cdot \widehat{\text{SNR}}}(\cdot)$ and $\text{M}^{\Lambda'}_{|A_j(\mathcal{N}_j)|^2 \cdot \widehat{\text{SNR}}}$, in function of $|A_j(\mathcal{N}_j)|^2 \cdot \widehat{\text{SNR}}$, $T_i(\mathcal{N}_i) \in \mathcal{CN}(0, \sigma^2_{T_i(\mathcal{N}_i)})$.

The $p_{err}^{\Lambda'}(A_j)$ is approximately coincidences with $p_{err}^{\Lambda_0}(A_j)$ because both operators find $l$ sub-channels however the threshold differs that leads to $p_{err}^{\Lambda'}(A_j) \geq p_{err}^{\Lambda_0}(A_j)$. In comparison to $\Lambda$, $p_{err}^{\Lambda}(A_j) \geq p_{err}^{\Lambda'}(A_j)$ because operator $\Lambda'$ always determines the required number $l$ of the sub-channels at a given threshold. It leads to the final conclusion $p_{err}^{\Lambda_0}(A_j) \leq p_{err}^{\Lambda'}(A_j) \leq p_{err}^{\Lambda}(A_j)$.

∎

# 4 Random Matrix Formalism of Multicarrier CVQKD

First, we summarize the basic functions and random matrix tools in Propositions 3 and 4, to the evaluate the results in Theorem 3 and Lemma 1. The proofs throughout Section 4 follow the notations of [27] and [28]. For the detailed description of the AMQD-MQA multiple-access multicarrier CKQKD scheme, see [3].

## 4.1 Multiuser Quadrature Allocation for Multicarrier CVQKD

**Proposition 3** (Random matrix decomposition of $F(\mathbf{T}(\mathcal{N}))$ of $K$ users.) *At $l$ Gaussian sub-channels and $K$ users, the $l \times K$ channel matrix $F(\mathbf{T}(\mathcal{N}))$ of $K$ users can be decomposed as $F(\mathbf{T}(\mathcal{N})) = \mathrm{XZ}\Phi$, where $\mathrm{Z}$ is an $l \times K$ random matrix, $\mathrm{X}$ is an $l \times l$ random matrix, while $\Phi$ is a $K \times K$ random matrix.*

*Proof.*
Let the $\mathrm{Z}_{i,j}$ entries of $\mathrm{Z}$ be independent, arbitrarily distributed complex random variables with identical mean $\mu_{i,j} = \mu$ and variance $\sigma_{i,j}^2 = 1/l$ such that the Lindeberg condition [27]

$$\frac{1}{K}\sum_{i,j} \int_{\{|\mathrm{Z}_{i,j}|\geq \delta\}} |\mathrm{Z}_{i,j}|^2 d\mathcal{P} = 0, \tag{107}$$

is satisfied, where $\mathcal{P}$ is a probability measure function that returns an event's probability.
Let $\mathrm{X}$ be an $l \times l$ random matrix, while $\Phi$ is a $K \times K$ random matrix, independent from each other and from $\mathrm{Z}$.
The $\eta$-transform of a nonnegative random variable $X$ is as

$$\eta_X(\gamma) = \mathbb{E}\left[\frac{1}{1+\gamma X}\right], \tag{108}$$

where $\gamma$ is a nonnegative real number, and $0 < \eta_X(\gamma) \leq 1$.
Without loss of generality, let the logical channel $\mathcal{M}_k$ of user $U_k$, defined as

$$\mathcal{M}_k = \{\mathcal{N}_0,...\mathcal{N}_{m-1}\}. \tag{109}$$

The single-carrier transmittance coefficient of $\mathcal{M}_k$ is referred to as $A_j(\mathcal{M}_k)$.
Particularly, then, for $K,l \to \infty$ and $K/l \to \chi$, where $\chi$ is a nonnegative variable, for the $\eta$-transform of $F(\mathbf{T}(\mathcal{N}))F(\mathbf{T}(\mathcal{N}))^\dagger$

$$\eta_{F(\mathbf{T}(\mathcal{N}))F(\mathbf{T}(\mathcal{N}))^\dagger} = \mathbb{E}\left[V_{F(\mathbf{T}(\mathcal{N}))F(\mathbf{T}(\mathcal{N}))^\dagger}(D,\gamma)\right], \tag{110}$$

where $D \in \mathcal{D}_{\{XX^\dagger\}}$ is an independent random variable with a distribution $\mathcal{D}_{\{XX^\dagger\}}$ of the asymptotic spectra of $XX^\dagger$, and

$$V_{F(\mathbf{T}(\mathcal{N}))F(\mathbf{T}(\mathcal{N}))^\dagger}(D,\gamma) = \frac{1}{1+\gamma\chi D\mathbb{E}\left[\frac{T}{1+\gamma T\mathbb{E}\left[DV_{F(\mathbf{T}(\mathcal{N}))F(\mathbf{T}(\mathcal{N}))^\dagger}(D,\gamma)\right]}\right]}, \tag{111}$$

where $T \in \mathcal{D}_{\{\Phi\Phi^\dagger\}}$ is an independent random variable with a distribution of the asymptotic spectra of $\Phi\Phi^\dagger$ [27].

The $\nu$-transform (Shannon transform) of a nonnegative random variable $X$ is defined as

$$\nu_X(\gamma) = \mathbb{E}\left[\log_2(1+\gamma X)\right], \tag{112}$$

where $\gamma$ is a nonnegative real number.

Let the $\nu$-transform of $F(\mathbf{T}(\mathcal{N}))F(\mathbf{T}(\mathcal{N}))^\dagger$ is as

$$\nu_{F(\mathbf{T}(\mathcal{N}))F(\mathbf{T}(\mathcal{N}))^\dagger}(\gamma) = \nu_{XX^\dagger}(\chi\gamma_d) + \chi\nu_{\Phi\Phi^\dagger}(\gamma_t) - \chi\frac{\gamma_d\gamma_t}{\gamma}\log_2 e, \tag{113}$$

where $\gamma_d$ and $\gamma_t$ are random variables such that [27]

$$\frac{\gamma_d\gamma_t}{\gamma} = 1 - \eta_{\Phi\Phi^\dagger}(\gamma_t) \tag{114}$$

and

$$\chi\frac{\gamma_d\gamma_t}{\gamma} = 1 - \eta_{XX^\dagger}(\chi\gamma_d). \tag{115}$$

Note that by denoting $\gamma_d(\gamma)$ the solution for (114) and (115), the result in (110) can be rewritten as

$$\eta_{F(\mathbf{T}(\mathcal{N}))F(\mathbf{T}(\mathcal{N}))^\dagger} = \eta_{XX^\dagger}(\chi\gamma_d(\gamma)). \tag{116}$$

Throughout the manuscript, let $X$ and $Y$ be defined as independent random variables $X \in \mathcal{U}_{[0,1]}$ and $Y \in \mathcal{U}_{[0,1]}$ drawn from a $\mathcal{U}$ uniform distribution on $[0,1]$.

Specifically, the decomposition of $F(\mathbf{T}(\mathcal{N})) = XZ\Phi$ with a $j$-th column $F(\mathbf{T}(\mathcal{N}))_j$ allows us to define a function $\mathrm{H}^{(l)}(a,\gamma)$, $(j-1)/K \leq a < j/K$, where $a$ is a random variable, precisely as

$$\mathrm{H}^{(l)}(a,\gamma) = \frac{1}{|F(\mathbf{T}(\mathcal{N}))_j|^2} F(\mathbf{T}(\mathcal{N}))_j^\dagger \left[I + \gamma\sum_{k\neq j} F(\mathbf{T}(\mathcal{N}))_k F(\mathbf{T}(\mathcal{N}))_k^\dagger\right]^{-1} F(\mathbf{T}(\mathcal{N}))_j; \tag{117}$$

where $I$ is the identity, such that for $K, l \to \infty$, and $a \in [0,1]$

$$\mathrm{H}^{(l)}(a,\gamma) \to \frac{\gamma_t(\gamma)}{\gamma\mathbb{E}[D]}; \tag{118}$$

where for $\gamma_t(\gamma)$ the result of (114) and (115) holds, and $\mathrm{H}^{(l)}$ converges to

$$\frac{\mathrm{H}(a,\gamma)}{\mathbb{E}[B(X,a)]}, \tag{119}$$

where $B(\cdot)$ is a limiting bounded measureable function [27], and $\mathrm{H}(a,\gamma)$ is evaluated as

$$\mathrm{H}(a,\gamma) = \mathbb{E}\left[B(X,a)\frac{1}{1+\gamma\chi\mathbb{E}\left[B(X,Y)\frac{1}{1+\gamma\mathrm{H}(Y,\gamma)}\big|X\right]}\right]. \tag{120}$$

Without loss of generality, the entries of $\mathrm{XZ}\Phi$ are independent, arbitrarily distributed zero-mean ($\mu_{i,j} = 0$) complex random variables with variance

$$\sigma_{i,j}^2 = \frac{\Pi_{i,j}}{l}, \tag{121}$$

where $\Pi$ is an $l \times K$ deterministic matrix with uniformly bounded $(i,j)$ entries, with variance profile $\mathrm{var}(a,b)$ [27]. Let $\mathrm{var}^l(\cdot)$ be the variance profile function as

$$\mathrm{var}^l : [0,1) \times [0,1) \to \mathbb{R} \tag{122}$$

such that for $\frac{i-1}{l} \leq a < \frac{i}{l}$ and $\frac{j-1}{K} \leq b < \frac{j}{K}$,

$$\mathrm{var}^l(a,b) = \Pi_{i,j}. \tag{123}$$

Particularly, if $\mathrm{var}^l(a,b)$ converges uniformly to a limiting bounded measurable function $B(a,b) = \mathrm{var}(a,b)$, then the limit

$$\mathrm{var}^l(a,b) \to B(a,b) = \mathrm{var}(a,b) \tag{124}$$

is the asymptotic variance profile [27] of $F(\mathbf{T}(\mathcal{N}))$. In this case (120) can be expressed as

$$\mathrm{H}(a,\gamma) = \mathbb{E}\left[\mathrm{var}(X,a)\frac{1}{1+\gamma\chi\mathbb{E}\left[\mathrm{var}(X,Y)\frac{1}{1+\gamma\mathrm{H}(Y,\gamma)}\big|X\right]}\right]. \tag{125}$$

Then, let us presume that for these entries the condition of

$$\frac{1}{K}\sum_{i,j}\int_{\{|F(\mathbf{T}(\mathcal{N}))_{i,j}|\geq\delta\}}\left|F(\mathbf{T}(\mathcal{N}))_{i,j}\right|^2 d\mathcal{P} = 0 \tag{126}$$

is satisfied [27], where $F(\mathbf{T}(\mathcal{N}))_{i,j}$ stands for the $(i,j)$-th entry of $F(\mathbf{T}(\mathcal{N}))$.

Precisely, if for the entries of $F(\mathbf{T}(\mathcal{N}))$ the conditions (121) and (126) are satisfied, then for $K, l \to \infty$, and $K/l \to \chi$, then for independent random variables $X \in \mathcal{U}_{[0,1]}$ and $Y \in \mathcal{U}_{[0,1]}$ drawn from a $\mathcal{U}$ uniform distribution on $[0,1]$,

$$\eta_{F(\mathbf{T}(\mathcal{N}))F(\mathbf{T}(\mathcal{N}))^\dagger}(\gamma) = \mathbb{E}\left[W_{F(\mathbf{T}(\mathcal{N}))F(\mathbf{T}(\mathcal{N}))^\dagger}(X,\gamma)\right], \tag{127}$$

where

$$W_{F(\mathbf{T}(\mathcal{N}))F(\mathbf{T}(\mathcal{N}))^\dagger}(a,\gamma) = \frac{1}{1+\chi\gamma\mathbb{E}\left[\mathrm{var}(a,Y)\Upsilon_{F(\mathbf{T}(\mathcal{N}))F(\mathbf{T}(\mathcal{N}))^\dagger}(Y,\gamma)\right]}, \tag{128}$$

and

$$\Upsilon_{F(\mathbf{T}(\mathcal{N}))F(\mathbf{T}(\mathcal{N}))^\dagger}(b,\gamma) = \frac{1}{1+\gamma\mathbb{E}\left[\mathrm{var}(X,b)\Gamma_{F(\mathbf{T}(\mathcal{N}))F(\mathbf{T}(\mathcal{N}))^\dagger}(X,\gamma)\right]}. \tag{129}$$

Note that it is precisely a coincidence with the $\eta$-transform of the empirical (squared) eigenvalue distribution of $F(\mathbf{T}(\mathcal{N}))F(\mathbf{T}(\mathcal{N}))^\dagger$ at $K,l \to \infty$ and $K/l \to \chi$ [27].

In particular, if for the entries of $F(\mathbf{T}(\mathcal{N}))$ the conditions (121) and (126) are satisfied, and $W_{F(\mathbf{T}(\mathcal{N}))F(\mathbf{T}(\mathcal{N}))^\dagger}(\cdot,\cdot)$ and $\Upsilon_{F(\mathbf{T}(\mathcal{N}))F(\mathbf{T}(\mathcal{N}))^\dagger}(\cdot,\cdot)$ are given as (128) and (129), respectively, then the $\nu$-transform of $F(\mathbf{T}(\mathcal{N}))F(\mathbf{T}(\mathcal{N}))^\dagger$ can be rewritten as

$$\nu_{F(\mathbf{T}(\mathcal{N}))F(\mathbf{T}(\mathcal{N}))^\dagger}(\gamma) = \chi \mathbb{E}\left[\log_2\left(1 + \gamma \mathbb{E}\left[\mathrm{var}(X,Y) W_{F(\mathbf{T}(\mathcal{N}))F(\mathbf{T}(\mathcal{N}))^\dagger}(X,\gamma)\big|Y\right]\right)\right]$$
$$+ \mathbb{E}\left[\log_2\left(1 + \gamma\chi \mathbb{E}\left[\mathrm{var}(X,Y) \Upsilon_{F(\mathbf{T}(\mathcal{N}))F(\mathbf{T}(\mathcal{N}))^\dagger}(Y,\gamma)\big|X\right]\right)\right]$$
$$- \gamma\chi \mathbb{E}\left[\mathrm{var}(X,Y) \Gamma_{F(\mathbf{T}(\mathcal{N}))F(\mathbf{T}(\mathcal{N}))^\dagger}(X,\gamma) \Upsilon_{F(\mathbf{T}(\mathcal{N}))F(\mathbf{T}(\mathcal{N}))^\dagger}(Y,\gamma)\right] \log_2 e. \tag{130}$$

Specifically, further note that for

$$\chi' = \chi \frac{\mathcal{P}(\mathbb{E}[\mathrm{var}(X,Y)|Y]\neq 0)}{\mathcal{P}(\mathbb{E}[\mathrm{var}(X,Y)|X]\neq 0)}, \tag{131}$$

the following limit holds,

$$\tau_\infty = \lim_{\gamma \to \infty}\left(\log_2(\gamma\chi) - \frac{\nu_{F(\mathbf{T}(\mathcal{N}))F(\mathbf{T}(\mathcal{N}))^\dagger}(\gamma)}{\min(\chi\mathcal{P}(\mathbb{E}[\mathrm{var}(X,Y)|Y]\neq 0), \mathcal{P}(\mathbb{E}[\mathrm{var}(X,Y)|X]\neq 0))}\right), \tag{132}$$

where $\tau_\infty$ depends on $\chi'$. Precisely $\tau_\infty$ differs for $\chi' < 1, \chi' = 1$ and $\chi' > 1$, as

$$\tau_{\infty,\chi'<1} = -\mathbb{E}\left[\log_2 \frac{\Gamma_\infty(Y')}{e} - \frac{1}{\chi'}\mathbb{E}\left[\log_2\left(1 + \mathbb{E}\left[\frac{\mathrm{var}(X',Y')}{W_\infty(Y')}\bigg|X'\right]\right)\right]\right], \tag{133}$$

where $W_\infty(\cdot)$ is yielded from

$$\mathbb{E}\left[\frac{1}{1+\mathbb{E}\left[\frac{\mathrm{var}(X',Y')}{W_\infty(Y')}\big|X'\right]}\right] = 1 - \chi', \tag{134}$$

while

$$\tau_{\infty,\chi'=1} = -\mathbb{E}\left[\log_2 \frac{\mathrm{var}(X',Y')}{e}\right] \tag{135}$$

and

$$\tau_{\infty,\chi'>1} = -\mathbb{E}\left[\log_2\left(\frac{1}{e}\mathbb{E}\left[\frac{\mathrm{var}(X',Y')}{1+\wp(Y')}\bigg|X'\right]\right)\right] - \chi'\mathbb{E}\left[\log_2(1 + \wp(Y'))\right], \tag{136}$$

where $\wp(\cdot)$ is yielded from

$$\wp(b) = \frac{1}{\chi'}\mathbb{E}\left[\frac{\mathrm{var}(X',b)}{\mathbb{E}\left[\frac{\mathrm{var}(X',Y')}{1+\wp(Y')}\big|X'\right]}\right]. \tag{137}$$

Precisely, the asymptotic theory of singular values of rectangular matrixes assumes the existence of an $l \times K$ matrix $\mathbf{M}$, for which the aspect ratio converges to a nonnegative variable [27] $\chi$,

$$\frac{K}{l} \to \chi, \tag{138}$$

as $K,l \to \infty$.

∎

**Proposition 4** (Channel profile of the $l \times K$ channel matrix $F(\mathbf{T}(\mathcal{N})) = \mathrm{X}\mathrm{Z}\Phi$ of $K$ users.) *The $\rho_{F(\mathbf{T}(\mathcal{N}))}(a,b) : [0,1]^2 \to \mathbb{R}$ function is the channel profile of $F(\mathbf{T}(\mathcal{N}))$, if for $X \in \mathcal{U}_{[0,1]}$ and $Y \in \mathcal{U}_{[0,1]}$, $\mathcal{D}_{\rho_{F(\mathbf{T}(\mathcal{N}))}(X,b)} = \mathcal{D}_{F_b}(\cdot)$ and $\mathcal{D}_{\rho_{F(\mathbf{T}(\mathcal{N}))}(a,Y)} = \mathcal{D}_{F_a}(\cdot)$, where $\mathcal{D}_{\rho_{F(\mathbf{T}(\mathcal{N}))}(X,b)}$ is the distribution of $\rho_{F(\mathbf{T}(\mathcal{N}))}(a,b)$, while $F_a(\cdot), F_b(\cdot)$ are nonrandom limits.*

Without loss of generality, let the $l \times K$ channel matrix of $K$ users given as $F(\mathbf{T}(\mathcal{N}))$. Focusing on the case that that Z has arbitrarily distributed $\mathrm{Z}_{i,j}$ zero-mean complex random variables with $\sigma^2_{\mathrm{Z}_{i,j}} = 1/l$, for a given $\mho$, $\mho \geq 0$ there exist a channel profile [27] function for $F(\mathbf{T}(\mathcal{N}))$ as

$$\rho_{F(\mathbf{T}(\mathcal{N}))}(a,b) : [0,1]^2 \to \mathbb{R}, \tag{139}$$

such that if $X \in \mathcal{U}_{[0,1]}$, then without loss of generality, the following relation holds for the $\mathcal{D}$ distribution of $\rho_{F(\mathbf{T}(\mathcal{N}))}(a,b)$:

$$\mathcal{D}_{\rho_{F(\mathbf{T}(\mathcal{N}))}(X,b)} = F_b(\cdot), \tag{140}$$

while if $Y \in \mathcal{U}_{[0,1]}$, then

$$\mathcal{D}_{\rho_{F(\mathbf{T}(\mathcal{N}))}(a,Y)} = F_a(\cdot), \tag{141}$$

where the nonrandom limits $F_a(\cdot), F_b(\cdot)$ for a given $a,b \in [0,1)$ have all bounded moments.

Then it can be shown that, if $F(\mathbf{T}(\mathcal{N})) = \mathrm{Z} \circ \mho$, where $\circ$ stands for the Hadamard product, then for $K, l \to \infty$, and $K/l \to \chi$, then for independent random variables $X \in \mathcal{U}_{[0,1]}$ and $Y \in \mathcal{U}_{[0,1]}$ the following coincidence holds:

$$\rho_{F(\mathbf{T}(\mathcal{N}))}(X,Y) = \mathrm{var}(X,Y). \tag{142}$$

**Theorem 3** (Random matrix formalism of multiple-access multicarrier CVQKD). *The random matrix decomposition of the $l \times K$ channel matrix $F(\mathbf{T}(\mathcal{N}))$ of the $K$ users is $F(\mathbf{T}(\mathcal{N})) = \mathrm{X} \circ \mathrm{Z}\Phi$, where X is an $l \times K$ random matrix $\mathrm{X} = (\mathcal{X}_0, \ldots, \mathcal{X}_{K-1})$, with $\mathcal{X}_i = \left[ A^{(k)}_{j,0}, \ldots, A^{(k)}_{j,l-1} \right]^T$ for user $U_k$, where $A^{(k)}_j = \frac{1}{m} \sum_{i=0}^{m-1} F(T_i(\mathcal{N}_i))$ is the averaged transmittance coefficients of $U_k$ derived by operator $\Lambda_0, \Lambda$, or $\Lambda'$, $\mathrm{Z} = (\Theta_0, \ldots, \Theta_{K-1})$ is an $l \times K$ random matrix, $\Theta_k = \left[ \odot_{k,0}, \ldots, \odot_{k,l-1} \right]^T$ is user $U_k$'s entry, $m$ is the number of sub-channels of $U_k$, $\circ$ is the Hadamard product, while $\Phi$ is an $l \times K$ random matrix, $\Phi = \mathrm{diag}(\Phi_0, \ldots, \Phi_{K-1})$.*

*Proof.*

Let the number of the users be $k \leq K$. The transmission of the users is realized through $l$ sub-channels, such that for each $U_k$ user, a given number $(m)$ of sub-channels are allocated. The transmittance coefficient statistics of the users are determined via operators $\Lambda_0, \Lambda, \Lambda'$, see Theorem 1. The derivation also utilizes the random matrix formulas introduced in Proposition 2 and 3, respectively. The proof utilizes the terms and notations of [27].

Let the $l \times K$ channel matrix $F(\mathbf{T}(\mathcal{N}))$ of $K$ users be given as

$$F(\mathbf{T}(\mathcal{N})) = \mathrm{X} \circ \mathrm{Z}\Phi, \tag{143}$$

where $\circ$ is the Hadamard (element wise) product, and

$$\mathrm{Z} = (\Theta_0, ..., \Theta_{K-1}) \tag{144}$$

is an $l \times K$ random matrix with zero-mean and $\sigma_Z^2 = 1/l$ variance entries, where

$$\Theta_k = [\odot_{k,0}, ..., \odot_{k,l-1}]^T \tag{145}$$

is user $U_k$'s entry, $\mathrm{X}$ is an $l \times K$ random matrix with independent entries for the users, such that

$$\mathrm{X} = (\mathcal{X}_0, ..., \mathcal{X}_{K-1}), \tag{146}$$

where for user $U_k$

$$\mathcal{X}_i = \left[A_{j,0}^{(k)}, ..., A_{j,l-1}^{(k)}\right]^T \tag{147}$$

where

$$A_j^{(k)} = \tfrac{1}{m}\sum_{i=0}^{m-1} F(T_i(\mathcal{N}_i)) \tag{148}$$

is the averaged transmittance coefficients of $U_k$ derived via a corresponding distribution statistics operator (see Theorems 1 and 2), $m$ is the number of sub-channels of $U_k$, while $\Phi$ is a $l \times K$ random matrix,

$$\Phi = diag(\Phi_0, ..., \Phi_{K-1}). \tag{149}$$

Specifically, for user $U_k$, let us identify $\mathcal{M}_k$ the logical channel (a set of $m$ sub-channels) of $U_k$ of the $k$-th column of $F(\mathbf{T}(\mathcal{N})) = (\mathcal{M}_0, ..., \mathcal{M}_{K-1})^T$ such that

$$\mathcal{M}_k = \left(\mathcal{M}_k^{(0)}, ..., \mathcal{M}_k^{(l)}\right)^T, \tag{150}$$

where $\mathcal{M}_k^{(i)}$ identifies the $i$-th logical sub-channel of $U_k$ such that

$$\mathcal{M}_k^{(i)} = \Phi_k A_{j,i}^{(k)} \odot_{k,i}. \tag{151}$$

Let $r$ be a nonnegative variable that identify the ratio of the sub-channels allocated to the logical channel $\mathcal{M}_k$ of $U_k$, as

$$r = \tfrac{m}{l}, \tag{152}$$

where $0 \leq r \leq 1$.

Using (143), the $\mathbf{Y}$ output matrix of the $K$ users can be written as

$$\begin{aligned}\mathbf{Y} &= F(\mathbf{T}(\mathcal{N}))\mathbf{X} + F(\Delta^{(K)}) \\ &= (\mathrm{X} \circ \mathrm{Z}\Phi)\mathbf{X} + F(\Delta^{(K)}),\end{aligned} \quad (153)$$

where $\mathbf{X}$ refers to the input matrix of the $K$ users, while $F(\Delta^{(K)})$ is the Fourier transformed matrix of the $\Delta^{(K)}$ noise matrix of the $K$ users.

Let us to utilize the results of Propositions 3 and 4, for $K, l \to \infty$ and $K/l \to \chi$, where $\chi$ is a nonnegative variable. Let $\eta_{U_k}$ be transmission efficiency of $U_k$ in function of the $\mathrm{SNR}^*_{U_k}$ (evaluated for the transmission of private classical information, see [5]) is as

$$\lim_{K \to \infty} \eta_{U_k}\left(\mathrm{SNR}^*_{U_k}\right) = \frac{\mathcal{F}\left(b, \mathrm{SNR}^*_{U_k}\right)}{\mathbb{E}[\rho_\vartheta(X,b)]}, \quad (154)$$

where $\rho_\vartheta(X,b)$ is the two-dimensional profile function of the $l \times K$ matrix $\vartheta$ and where an $(r,t)$ entry of $\vartheta$ is

$$\vartheta_{r,t} = A^{(t)}_{j,r} \Phi_t, \quad (155)$$

which identifies the $r$-th logical sub-channel of the $t$-th user [27], and let

$$b = \frac{k}{K}, \quad (156)$$

where $0 \leq b \leq 1$, and $X \in \mathcal{U}_{[0,1]}$ and $Y \in \mathcal{U}_{[0,1]}$.

The $\mathrm{SNR}^*_{U_k}$ of $U_k$ is as

$$\mathrm{SNR}^*_{U_k} = \sum_{i=0}^{m-1} \mathrm{SNR}^*_i = \sum_{i=0}^{m-1} \frac{\sigma^2_{\omega_i}}{\sigma^2_{\mathcal{N}^*_i}}, \quad (157)$$

where $m$ is the number of Gaussian subcarrier CVs allocated to $U_k$, $\sigma^2_{\omega_i}$ is the constant subcarrier modulation variance, while $\sigma^2_{\mathcal{N}^*_i}$ is expressed precisely as

$$\sigma^2_{\mathcal{N}^*_i} = \sigma^2_{\omega_i}\left(\frac{\sigma^2_{\omega_i}|F(T_i(\mathcal{N}_i))|^2 + \sigma^2_{\mathrm{X}_i}}{1 + \sigma^2_{\mathrm{X}_i}\sigma^2_{\omega_i}|F(T_i(\mathcal{N}_i))|^2} - 1\right)^{-1}, \quad (158)$$

where

$$\sigma^2_{\mathrm{X}_i} = \sigma^2_0 + N_i, \quad (159)$$

where $\sigma^2_0$ is the vacuum noise and $N_i$ is the excess noise of the Gaussian sub-channel $\mathcal{N}_i$ defined as

$$N_i = \frac{(W_i - 1)\left(|F(T_{Eve,i})|^2\right)}{1 - |F(T_{Eve,i})|^2}, \quad (160)$$

where $W_i$ is the variance of Eve's EPR state used for the attacking of $\mathcal{N}_i$, while

$$|T_{Eve,i}|^2 = 1 - |T_i|^2 \quad (161)$$

is the transmittance of Eve's beam splitter (BS), while $|T_i|^2$ is the transmittance of $\mathcal{N}_i$, while function $\mathcal{F}\left(b, \mathrm{SNR}^*_{U_k}\right)$ is yielded as

$$\mathcal{F}\left(b,\mathrm{SNR}_{U_k}^*\right) = \mathbb{E}\left[\frac{\rho_\vartheta(X,b)}{1+\mathrm{SNR}_{U_k}^*\chi\mathbb{E}\left[\frac{\rho_\vartheta(X,Y)}{1+\mathrm{SNR}_{U_k}^*\mathcal{F}(Y,\mathrm{SNR}_{U_k}^*)}\Big|X\right]}\right]. \tag{162}$$

Particularly, the result in (162) can be rewritten as

$$\mathcal{F}\left(b,\mathrm{SNR}_{U_k}^*\right) = \mathbb{E}\left[\rho_\vartheta(X,b)\psi\left(X,\mathrm{SNR}_{U_k}^*\right)\right], \tag{163}$$

where function $\psi\left(X,\mathrm{SNR}_{U_k}^*\right)$ is as

$$\psi\left(X,\mathrm{SNR}_{U_k}^*\right) = \frac{1}{1+\mathrm{SNR}_{U_k}^*\mathbb{E}\left[\frac{\rho_\vartheta(X,Y)}{1+\mathrm{SNR}_{U_k}^*\mathbb{E}\left[\psi(X,\mathrm{SNR}_{U_k}^*)\rho_\vartheta(X,Y)|Y\right]}\right]}. \tag{164}$$

Then let us assume that for $\rho_\vartheta(a,b)$ the following relation holds [27]:

$$\rho_\vartheta(a,b) \approx |\Phi_k|^2 \left|A_{j,i}^{(k)}\right|^2, \tag{165}$$

where $(i-1)/l \leq a < i/l$, and $(k-1)/K \leq b < k/K$. Specifically, in this case, $\eta_{U_k}\left(\mathrm{SNR}_{U_k}^*\right)$ is yielded as

$$\eta_{U_k}\left(\mathrm{SNR}_{U_k}^*\right) = \frac{\varphi_k\left(\mathrm{SNR}_{U_k}^*\right)}{\left|A_{j,i}^{(k)}\right|^2} = \frac{\varphi_k\left(\mathrm{SNR}_{U_k}^*\right)}{\frac{1}{m}\sum_{i=0}^{m-1}F(T_i(\mathcal{N}_i))}, \tag{166}$$

where $m$ identifies the number of sub-channels of $U_k$ used in the transmission, that is, for remaining $l-k$ sub-channels $\left|F(T_i(\mathcal{N}_i))\right|^2 = 0$, and $\varphi_k\left(\mathrm{SNR}_{U_k}^*\right)$ is expressed as

$$\varphi_k\left(\mathrm{SNR}_{U_k}^*\right) = \frac{1}{l}\sum_{r=0}^{l-1}\frac{\left|A_{r,k}^{(k)}\right|^2}{1+\mathrm{SNR}_{U_k}^*\frac{\chi}{K}\sum_{h=0}^{K-1}\frac{|\Phi_h|^2\left|A_{r,h}^{(h)}\right|^2}{1+\mathrm{SNR}_{U_k}^*|\Phi_h|^2\varphi_h(\mathrm{SNR}_{U_h}^*)}} = \frac{1}{l}\sum_{r=0}^{l-1}\frac{\frac{1}{m}\sum_{i=0}^{m-1}F(T_i(\mathcal{N}_i))}{1+\mathrm{SNR}_{U_k}^*\frac{\chi}{K}\sum_{h=0}^{K-1}\frac{|\Phi_h|^2\left|A_{r,h}^{(h)}\right|^2}{1+\mathrm{SNR}_{U_h}^*\left|A_{r,k}^{(k)}\right|^2\varphi_h(\mathrm{SNR}_{U_h}^*)}}, \tag{167}$$

where $(r,k)$ identifies the $r$-th subcarrier of user $U_k$.

Using (166), an $f_{\mathit{eff}}(\cdot)$ efficiency function is introduced as

$$f_{\mathit{eff}}\left(\chi,\mathrm{SNR}_{U_k}^*\right) = \chi\mathbb{E}\left[\log_2\left(1+\mathrm{SNR}_{U_k}^*\mathcal{F}\left(Y,\mathrm{SNR}_{U_k}^*\right)\right)\right], \tag{168}$$

where $\mathcal{F}(\cdot)$ is derived via (162).

Using Proposition 2, let

$$\chi\frac{\mathcal{P}(\mathbb{E}[\rho_\vartheta(X,Y)|Y]>0)}{\mathcal{P}(\mathbb{E}[\rho_\vartheta(X,Y)|X]>0)} < 1. \tag{169}$$

If (169) holds, then without loss of generality,

$$f_{\mathit{eff}}\left(\chi,\mathrm{SNR}_{U_k}^*\right) = \chi\mathbb{E}\left[\log_2\left(1+\mathrm{SNR}_{U_k}^*\mathcal{F}\left(Y,\mathrm{SNR}_{U_k}^*\right)\right)\right], \tag{170}$$

and if $\mathrm{SNR}_{U_k}^* \to \infty$ also holds, then

$$\mathcal{F}\left(b, \text{SNR}^*_{U_k}\right) \to \mathbb{E}\left[\frac{\rho_\vartheta(X,b)}{1+\chi\mathbb{E}\left[\frac{\rho_\vartheta(X,Y)}{\mathcal{F}(Y)}\big|X\right]}\right]. \tag{171}$$

Specifically, using (168), the $P_{sym}(\mathcal{M}_k)$ symmetric private classical capacity (the maximum common rate at which the $K$ users both can reliably transmit private classical information over the $l$ sub-channels of $\mathcal{N}$) of the $\mathcal{M}_k$ logical channel of $U_k$ is expressed as follows

$$\begin{aligned}
P_{sym}(\mathcal{M}_k) &= \tfrac{1}{K}\Big(f_{eff}\Big(\chi, \textstyle\sum_{i=0}^{l-1}\text{SNR}^*_i\Big) \\
&\quad + \mathbb{E}\Big[\log_2\Big(1 + \textstyle\sum_{i=0}^{l-1}\text{SNR}^*_i \chi \mathbb{E}\Big[\rho_\vartheta(X,Y)\Upsilon\Big(Y, \textstyle\sum_{i=0}^{l-1}\text{SNR}^*_i\Big)\Big|X\Big]\Big)\Big] \\
&\quad - \chi \textstyle\sum_{i=0}^{l-1}\text{SNR}^*_i \mathbb{E}\Big[\mathcal{F}\Big(Y, \textstyle\sum_{i=0}^{l-1}\text{SNR}^*_i\Big)\Upsilon\Big(Y, \textstyle\sum_{i=0}^{l-1}\text{SNR}^*_i\Big)\Big]\log_2 e\Big) \\
&= \tfrac{1}{K}\Big(f_{eff}\Big(\chi, \textstyle\sum_{i=0}^{l-1}\text{SNR}^*_i\Big) + \mathbb{E}\Big[\log_2\Big(\tfrac{1}{\psi(X,\sum_{i=0}^{l-1}\text{SNR}^*_i)}\Big)\Big] \\
&\quad + \Big(\mathbb{E}\Big[\psi\Big(X, \textstyle\sum_{i=0}^{l-1}\text{SNR}^*_i\Big)\Big] - 1\Big)\log_2 e\Big),
\end{aligned} \tag{172}$$

where $\sum_{i=0}^{l-1}\text{SNR}^*_i$ is the total SNRs of the $l$ Gaussian sub-channels evaluated for the transmission of private classical information expressed as $\sum_{i=0}^{l-1}\text{SNR}^*_i = \sum_{i=0}^{l-1}\sigma^2_{\omega_i}/\sigma^2_{\mathcal{N}^*_i}$, where $\sigma^2_{\mathcal{N}^*_i}$ is shown in (158), $X \in \mathcal{U}_{[0,1]}$ and $Y \in \mathcal{U}_{[0,1]}$, and

$$\mathcal{F}\Big(b, \textstyle\sum_{i=0}^{l-1}\text{SNR}^*_i\Big) = \mathbb{E}\left[\frac{\rho_\vartheta(X,b)}{1+\chi\sum_{i=0}^{l-1}\text{SNR}^*_i \mathbb{E}\left[\rho_\vartheta(X,Y)\Upsilon\left(Y,\sum_{i=0}^{l-1}\text{SNR}^*_i\right)\big|X\right]}\right], \tag{173}$$

and function $\Upsilon(\cdot)$ is as

$$\Upsilon\Big(b, \textstyle\sum_{i=0}^{l-1}\text{SNR}^*_i\Big) = \frac{1}{1+\sum_{i=0}^{l-1}\text{SNR}^*_i \mathcal{F}\left(b,\sum_{i=0}^{l-1}\text{SNR}^*_i\right)}. \tag{174}$$

In particular, defining function $\ell\Big(b, \sum_{i=0}^{l-1}\text{SNR}^*_i\Big)$ precisely as

$$\ell\Big(b, \textstyle\sum_{i=0}^{l-1}\text{SNR}^*_i\Big) = \mathbb{E}\left[\frac{\rho_\vartheta(X,b)}{1+\sum_{i=0}^{l-1}\text{SNR}^*_i\chi(1-b)\mathbb{E}\left[\rho_\vartheta(X,Z)\frac{1}{1+\sum_{i=0}^{l-1}\text{SNR}^*_i\ell\left(Z,\sum_{i=0}^{l-1}\text{SNR}^*_i\right)}\big|X\right]}\right], \tag{175}$$

where $Z \in \mathcal{U}_{[b,1]}$ is an independent random variable, drawn from a $\mathcal{U}$ uniform distribution on $[b,1]$, the formula of (172) can be rewritten in a simplified form specifically as

$$P_{sym}(\mathcal{M}_k) = \tfrac{1}{K}\chi\mathbb{E}\left[\log_2\Big(1 + \textstyle\sum_{i=0}^{l-1}\text{SNR}^*_i \ell\Big(Y, \textstyle\sum_{i=0}^{l-1}\text{SNR}^*_i\Big)\Big)\right]. \tag{176}$$

From (176), the $S_{sym}(U_k)$ symmetric secret key rate (a common rate at which the $K$ users both can reliably transmit private classical information over the $l$ sub-channels of $\mathcal{N}$) of $U_k$ is as

$$S_{sym}(U_k) \leq \tfrac{1}{K}\chi\mathbb{E}\left[\log_2\Big(1 + \textstyle\sum_{i=0}^{l-1}\text{SNR}^*_i \ell\Big(Y, \textstyle\sum_{i=0}^{l-1}\text{SNR}^*_i\Big)\Big)\right]. \tag{177}$$

The $P_{sym}(\mathcal{M}_k)$ of $U_k$ from (176) in function of $r$ is depicted in Fig. 5.

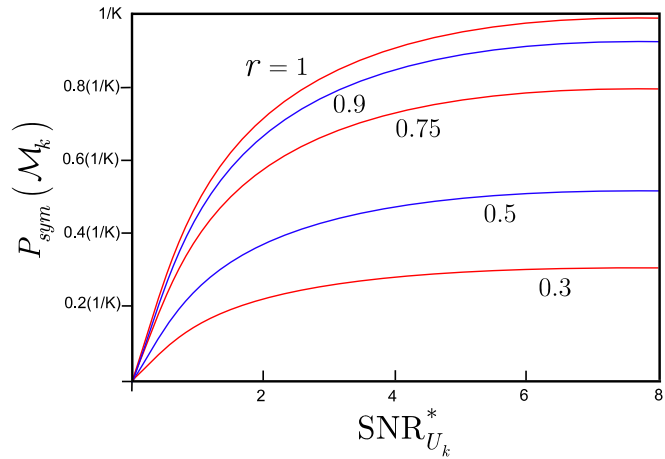

**Figure 5**. The $P_{sym}(\mathcal{M}_k)$ of $U_k$ in function of $\text{SNR}^*_{U_k}$ for $0 \leq r \leq 1$, where $\text{SNR}^*_{U_k}$ is the SNR for the transmission of private classical information over the $\mathcal{M}_k$ logical channel of $U_k$, via $m$ Gaussian sub-channels; $r = m/l$ identifies the ratio of the sub-channels allocated to $U_k$.

## 4.2 Multiuser Transmission over Identical Gaussian Sub-channels

**Lemma 1** (Random matrix formalism of multiple-access multicarrier CVQKD for identically allocated sub-channels). *Let set* $\mathcal{S} = \{\mathcal{N}_i\}, i = 0, \ldots, h-1, \; h \leq l$, *refer to the sub-channels of the* $1 \leq k \leq K \leq l$ *users, obtained via* $\Lambda_0, \Lambda, \Lambda'$. *The random matrix decomposition of the* $h \times K$ *channel matrix* $F(\mathbf{T}(\mathcal{N}))$ *is* $F(\mathbf{T}(\mathcal{N})) = \mathrm{X}\mathrm{U}\Phi$, *where* $\mathrm{U} = (\mathbf{u}_0, \ldots, \mathbf{u}_{K-1})$ *is an* $h \times K$ *isotropic unitary matrix,* $\mathbf{u}_k = [q_{k,0}, \ldots, q_{k,h-1}]^T$, $\mathrm{X}$ *is an* $h \times h$ *random matrix* $\mathrm{X} = diag(\mathcal{X}_0, \ldots, \mathcal{X}_{h-1})$, *with* $\mathcal{X}_k = \left[A^{(k)}_{j,0}, \ldots, A^{(k)}_{j,h-1}\right]^T$, *where* $A^{(k)}_j = \frac{1}{h}\sum_{i=0}^{h-1} F(T_i(\mathcal{N}_i))$ *is the averaged transmittance coefficients of* $U_k$ *in* $\mathcal{S}$, *such that* $\mathcal{X}_i = \mathcal{X}_k$ *for* $i = 0, \ldots, K-1$, *while* $\Phi$ *is a* $K \times K$ *random matrix,* $\Phi = (\Phi_0, \ldots, \Phi_{K-1})$.

*Proof.*
The proof utilizes Proposition 3. Let us identify set
$$\mathcal{S} = \{\mathcal{N}_i\}, i = 0, \ldots, h-1, \; h \leq l, \tag{178}$$
the sub-channels allocated to the $1 \leq k \leq K \leq l$ users. The transmittance coefficients are determined via operators $\Lambda_0, \Lambda, \Lambda'$ (see Theorem 1). The proof utilizes the terms and notations of [27].

The random matrix decomposition of the $h \times K$ channel matrix $F(\mathbf{T}(\mathcal{N}))$ is
$$F(\mathbf{T}(\mathcal{N})) = \mathrm{X}\mathrm{U}\Phi, \tag{179}$$
where
$$\mathrm{U} = (\mathbf{u}_0, \ldots, \mathbf{u}_{K-1}) \tag{180}$$

is an $h \times K$ isotropic unitary matrix, independent of X and $\Phi$, $UU^\dagger = I$, with arbitrarily distributed random variables such that the Lindeberg condition (see (107)) is satisfied, with mean $\mu_{i,j} = \mu$ and variance $\sigma^2_{i,j} = 1/l$,

$$\mathbf{u}_k = [q_{k,0},\ldots,q_{k,h-1}]^T, \tag{181}$$

X is an $h \times h$ random matrix

$$X = diag(\mathcal{X}_0,\ldots,\mathcal{X}_{h-1}), \tag{182}$$

where

$$\mathcal{X}_k = \left[A^{(k)}_{j,0},\ldots,A^{(k)}_{j,h-1}\right]^T, \tag{183}$$

and

$$A^{(k)}_j = \tfrac{1}{h}\sum_{i=0}^{h-1} F(T_i(\mathcal{N}_i)) \tag{184}$$

is the averaged transmittance coefficients of $U_k$ taken over $\mathcal{S}$, such that

$$\mathcal{X}_i = \mathcal{X}_k \text{ for } i = 0,\ldots,K-1, \tag{185}$$

while $\Phi$ is a $K \times K$ random matrix,

$$\Phi = (\Phi_0,\ldots,\Phi_{K-1}). \tag{186}$$

Precisely, if the users transmit through the same subcarriers, then for all $U_i, i = 0,\ldots,k-1$ users, an identical $\mathcal{M}_{U_i} = \mathcal{M}_k$ logical channel is allocated, see (185); thus (182) can be rewritten as [27]

$$X = diag(\mathcal{X}_k), \tag{187}$$

with the averaged coefficient of (184) for the $h$ sub-channels of user $U_k$.

Particularly, since all users use the same Gaussian sub-channels, the empirical distribution of $XX^\dagger$ converges to a nonrandom limit $F_{|C|^2}$, thus

$$\mathcal{D}_{XX^\dagger} \to F_{|C|^2}, \tag{188}$$

where $|C|$ is a random variable having a distribution of the asymptotic singular value distribution of the singular values of X, i.e., the distribution of $|C|^2$ is determined by the asymptotic spectra of $XX^\dagger$, thus without loss of generality

$$|C| = |A_j(\mathcal{M}_k)| = \left|A^{(k)}_j\right|, \tag{189}$$

where $A_j(\mathcal{M}_k)$ is the $A_j$ coefficient of the $\mathcal{M}_k$ channel of $U_k$, see (184). Let $|\mho|$ be a random variable having a distribution of the asymptotic singular value distribution of the singular values of $\Phi$, the distribution of $|\mho|^2$ is determined by the asymptotic spectra of $\Phi\Phi^\dagger$.

Specifically, the $f_{eff}(\cdot)$ efficiency function in (168) can be rewritten as

$$f_{eff}\left(\chi, \text{SNR}^*_{U_k}\right) = \chi \mathbb{E}\left[\log_2\left(1 + \mathbb{E}\left[\left|A_j(\mathcal{M}_k)\right|^2\right]\text{SNR}^*_{U_k}\eta_{U_k}\left(\text{SNR}^*_{U_k}\right)\right)\right], \tag{190}$$

where $\eta_{U_k}$ is the multiuser efficiency, evaluated as

$$\eta_{U_k} = \frac{1}{\mathbb{E}\left[|C|^2\right]} \mathbb{E}\left[\frac{|C|^2}{1+\text{SNR}_{U_k}^*\chi|C|^2\mathbb{E}\left[\frac{|\upsilon|^2}{1+\mathbb{E}\left[|C|^2\right]\text{SNR}_{U_k}^*|\upsilon|^2\eta_{U_k}}\right]}\right] \quad (191)$$

$$= \frac{1}{\mathbb{E}\left[|A_j(\mathcal{M}_k)|^2\right]} \mathbb{E}\left[\frac{|A_j(\mathcal{M}_k)|^2}{1+\text{SNR}_{U_k}^*\chi|A_j(\mathcal{M}_k)|^2\mathbb{E}\left[\frac{|\upsilon|^2}{1+\mathbb{E}\left[|A_j(\mathcal{M}_k)|^2\right]\text{SNR}_{U_k}^*|\upsilon|^2\eta_{U_k}}\right]}\right],$$

such that

$$\eta\left(\text{SNR}_{U_k}^*\right) \to \eta\left(\text{SNR}_{U_k}^*\mathbb{E}\left[|A_j(\mathcal{M}_k)|^2\right]\right), \quad (192)$$

where

$$\eta\left(\text{SNR}_{U_k}^*\mathbb{E}\left[|A_j(\mathcal{M}_k)|^2\right]\right) = \frac{\eta_{U_k}}{1+\left(\text{SNR}_{U_k}^*\mathbb{E}\left[|A_j(\mathcal{M}_k)|^2\right]\right)\eta_{U_k}}$$

$$= \mathbb{E}\left[\frac{|A_j(\mathcal{M}_k)|^2 / \mathbb{E}\left[|A_j(\mathcal{M}_k)|^2\right]}{\chi\left(\text{SNR}_{U_k}^*\mathbb{E}\left[|A_j(\mathcal{M}_k)|^2\right]\right)\left(|A_j(\mathcal{M}_k)|^2 / \mathbb{E}\left[|A_j(\mathcal{M}_k)|^2\right]\right)+1+(1-\chi)\left(\text{SNR}_{U_k}^*\mathbb{E}\left[|A_j(\mathcal{M}_k)|^2\right]\right)\eta_{U_k}}\right]. \quad (193)$$

Without loss of generality, the term $\eta_{U_k}\left(\text{SNR}_{U_k}^*\right)$ in (190) for $0 \leq \chi \leq 1$ is expressed as

$$\eta_{U_k}\left(\text{SNR}_{U_k}^*\right) = (1-\chi). \quad (194)$$

Precisely, the $P_{sym}(\mathcal{M}_k)$ symmetric private classical capacity of the $\mathcal{M}_k$ logical channel of $U_k$ is particularly yielded as

$$P_{sym}(\mathcal{M}_k) = \frac{1}{K}\chi\mathbb{E}\left[\log_2\left(1+\ell\left(Y,\sum_{i=0}^{l-1}\text{SNR}_i^*\right)\right)\right]$$
$$= \frac{1}{K}\int_0^{\sum_{i=0}^{l-1}\text{SNR}_i^*}\frac{1}{x}\left(1-\eta_{F(\mathbf{T}(\mathcal{N}))F(\mathbf{T}(\mathcal{N}))^\dagger}(x)\right)dx, \quad (195)$$

where $\sum_{i=0}^{l-1}\text{SNR}_i^* = \sum_l \sigma_{\omega_i}^2 / \sigma_{\mathcal{N}_i^*}^2$, $Y \in \mathcal{U}_{[0,1]}$, and $\eta_{F(\mathbf{T}(\mathcal{N}))F(\mathbf{T}(\mathcal{N}))^\dagger}$ is the $\eta$-transform of $F(\mathbf{T}(\mathcal{N}))F(\mathbf{T}(\mathcal{N}))^\dagger$, specifically as

$$\eta_{F(\mathbf{T}(\mathcal{N}))F(\mathbf{T}(\mathcal{N}))^\dagger}\left(\sum_{i=0}^{l-1}\text{SNR}_i^*\right) = \eta_{XX^\dagger}\left(\sum_{i=0}^{l-1}\text{SNR}_i^*\frac{\chi-1+\eta_{F(\mathbf{T}(\mathcal{N}))F(\mathbf{T}(\mathcal{N}))^\dagger}}{\eta_{F(\mathbf{T}(\mathcal{N}))F(\mathbf{T}(\mathcal{N}))^\dagger}\left(\sum_{i=0}^{l-1}\text{SNR}_i^*\right)}\right), \quad (196)$$

while function $\ell\left(Y,\sum_{i=0}^{l-1}\text{SNR}_i^*\right)$ [27] is as

$$\frac{\ell\left(Y,\sum_{i=0}^{l-1}\text{SNR}_i^*\right)}{1+\ell\left(Y,\sum_{i=0}^{l-1}\text{SNR}_i^*\right)} = \mathbb{E}\left[\frac{\sum_{i=0}^{l-1}\text{SNR}_i^*|F(T_i(\mathcal{N}_i))|^2}{\chi Y\sum_{i=0}^{l-1}\text{SNR}_i^*|F(T_i(\mathcal{N}_i))|^2+1+(1-\chi Y)\ell\left(Y,\sum_{i=0}^{l-1}\text{SNR}_i^*\right)}\right]. \quad (197)$$

Using (196), the following relation holds for the $K$ users

$$\frac{1}{K}Tr\left[\left(I+\sum_{i=0}^{l-1}\text{SNR}_i^*F(\mathbf{T}(\mathcal{N}))^\dagger F(\mathbf{T}(\mathcal{N}))\right)^{-1}\right]$$
$$= 1-\frac{1}{\chi}\left(1-\eta_{F(\mathbf{T}(\mathcal{N}))F(\mathbf{T}(\mathcal{N}))^\dagger}\left(\sum_{i=0}^{l-1}\text{SNR}_i^*\right)\right). \quad (198)$$

Precisely, the $S_{sym}(U_k)$ symmetric secret key rate of $U_k$ over identical sub-channels is expressed as

$$S_{sym}(U_k) \leq \tfrac{1}{K}\int_0^{\sum_{i=0}^{l-1}\mathrm{SNR}_i^*} \tfrac{1}{x}\bigg(1 - \eta_{F(\mathbf{T}(\mathcal{N}))F(\mathbf{T}(\mathcal{N}))^\dagger}(x)\bigg)dx. \tag{199}$$

In particular, using the density function $f_\chi(x)$ (Marchenko-Pastur density function [27]),

$$f_\chi(x) = \left(1 - \tfrac{1}{\chi}\right)^+ \xi(x) + \frac{\sqrt{(x-u)^+(v-x)^+}}{2\pi\chi x}, \tag{200}$$

where $\xi(\cdot)$ is a related random function, and $u = \left(1 - \sqrt{\chi}\right)^2$ and $v = \left(1 + \sqrt{\chi}\right)^2$, (198) can be rewritten precisely as

$$\begin{aligned}
&\tfrac{1}{K}Tr\left[\left(I + \sum_{i=0}^{l-1}\mathrm{SNR}_i^* F(\mathbf{T}(\mathcal{N}))^\dagger F(\mathbf{T}(\mathcal{N}))\right)^{-1}\right] \\
&= \int_a^b \frac{1}{1+\sum_{i=0}^{l-1}\mathrm{SNR}_i^* x} f_\chi(x)\,dx.
\end{aligned} \tag{201}$$

The density function $f_\chi(x)$ for $\chi = 0.2, 0.5$ and $\chi = 1$ is depicted in Fig. 6.

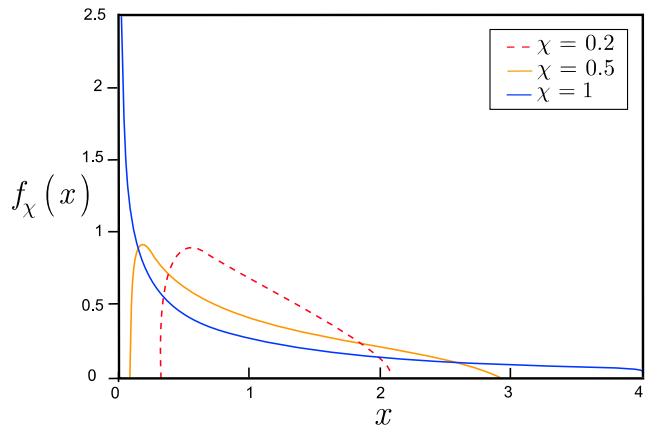

**Figure 6**. The density function $f_\chi(x)$ for $\chi = 0.2, 0.5, 1$.

∎

## 5 Conclusions

The multicarrier CVQKD systems are aimed to avoid the main drawbacks of CVQKD, such as low secret key rates, low tolerable losses and short communication distances, allowing the legal parties to establish an unconditional secure communication over the standard networks. We provided a distribution statistics and random matrix framework for multicarrier CVQKD. Exploiting order statistics, we defined different statistics methods to derive the averaged transmittance coefficients from the Gaussian sub-channel transmittances. The provided operators order, select, and sum the sub-channel coefficients for the information transmission at a given threshold. We analyzed the complexity and the error probabilities of the sub-channel selection operators and defined an optimized progressive sub-channel scanning scheme. We characterized a random matrix formalism for multicarrier CVQKD that utilizes the mathematical background of random matrix theory. The framework is formulated for multiple-access multicarrier CVQKD, through the multiuser quadrature allocation multiuser transmission scheme. Using the random matrix formalism, we studied the private classical information capabilities of the users at different subcarrier allocation mechanisms. The proposed combined framework of order statistics and random matrix theory is particularly convenient for the firm portrayal of the information flowing in experimental multicarrier CVQKD scenarios.

# Acknowledgements

The author would like to thank Professor Sandor Imre for useful discussions. This work was partially supported by the GOP-1.1.1-11-2012-0092 (*Secure quantum key distribution between two units on optical fiber network*) project sponsored by the EU and European Structural Fund, and by the COST Action MP1006.

# Supplemental Information

## S.1 Notations

The notations of the manuscript are summarized in Table S.1.

**Table S.1.** Summary of notations.

| | |
|---|---|
| $i$ | Index for the $i$-th subcarrier Gaussian CV, $\lvert\phi_i\rangle = x_i + \mathrm{i}p_i$. |
| $j$ | Index for the $j$-th Gaussian single-carrier CV, $\lvert\varphi_j\rangle = x_j + \mathrm{i}p_j$. |
| $l$ | Number of Gaussian sub-channels $\mathcal{N}_i$ for the transmission of the Gaussian subcarriers. The overall number of the sub-channels is $n$. The remaining $n-l$ sub-channels do not transmit valuable information. |
| $x_i, p_i$ | Position and momentum quadratures of the $i$-th Gaussian subcarrier, $\lvert\phi_i\rangle = x_i + \mathrm{i}p_i$. |
| $x'_i, p'_i$ | Noisy position and momentum quadratures of Bob's $i$-th noisy subcarrier Gaussian CV, $\lvert\phi'_i\rangle = x'_i + \mathrm{i}p'_i$. |
| $x_j, p_j$ | Position and momentum quadratures of the $j$-th Gaussian single-carrier $\lvert\varphi_j\rangle = x_j + \mathrm{i}p_j$. |
| $x'_j, p'_j$ | Noisy position and momentum quadratures of Bob's $j$-th recovered single-carrier Gaussian CV $\lvert\varphi'_j\rangle = x'_j + \mathrm{i}p'_j$. |
| $x_{A,i}, p_{A,i}$ | Alice's quadratures in the transmission of the $i$-th subcarrier. |
| $\lvert\phi_i\rangle, \lvert\phi'_i\rangle$ | Transmitted and received Gaussian subcarriers. |
| $Q(\cdot)$ | Gaussian tail function. |
| $\mathbf{z} \in \mathcal{CN}(0, \mathbf{K}_\mathbf{z})$ | A $d$-dimensional input CV vector to transmit valuable information. |
| $\mathbf{z}'^T$ | A $d$-dimensional noisy output vector, $\mathbf{z}'^T = \mathbf{A}^\dagger \mathbf{z} + \left(F^d(\Delta)\right)^T = (z'_0, \ldots, z'_{d-1})$, where $z'_j = \left(\frac{1}{l}\sum_{i=0}^{l-1} F(T_{j,i}(\mathcal{N}_{j,i}))\right) z_j + F(\Delta) \in \mathcal{CN}\left(0, 2\left(\sigma^2_{\omega_0} + \sigma^2_\mathcal{N}\right)\right)$. |
| $\mathrm{SNR}^*_{U_k}$ | The SNR quantity of $U_k$, evaluated for the transmission of private classical information. |
| $\sum_{i=0}^{l-1} \mathrm{SNR}^*_i$ | Total SNR of the $l$ Gaussian sub-channels, evaluated for the transmission of private classical information, $\sum_{i=0}^{l-1} \mathrm{SNR}^*_i = \sum_{i=0}^{l-1} \sigma^2_{\omega_i} / \sigma^2_{\mathcal{N}^*_i}$, where $\sigma^2_{\omega_i}$ is the constant subcarrier modulation variance, and $$\sigma^2_{\mathcal{N}^*_i} = \sigma^2_{\omega_i} \left(\frac{\sigma^2_{\omega_i}|F(T_i(\mathcal{N}_i))|^2 + \sigma^2_{X_i}}{1 + \sigma^2_{X_i}\sigma^2_{\omega_i}|F(T_i(\mathcal{N}_i))|^2} - 1\right)^{-1},$$ where $\sigma^2_{X_i} = \sigma^2_0 + N_i$, $\sigma^2_o$ is the vacuum noise, $N_i$ is the excess noise of the Gaussian sub-channel $\mathcal{N}_i$ as $$N_i = \frac{(W_i - 1)\left(|F(T_{Eve,i})|^2\right)}{1 - |F(T_{Eve,i})|^2},$$ where $W_i$ is the variance of Eve's EPR state used for the attacking of $\mathcal{N}_i$, while $|T_{Eve,i}|^2 = 1 - |T_i|^2$ is the transmittance of Eve's beam splitter (BS), and $|T_i|^2$ is the transmittance of $\mathcal{N}_i$. |
| $\frac{1}{l}\left|F(\widehat{T_i}(\mathcal{N}_i))\right|^2$ | Normalized, unordered sub-channel transmittance coefficient, real variable. |
| $\frac{1}{l}\left|F(T_i(\mathcal{N}_i))\right|^2$ | Normalized, ordered sub-channel transmittance coefficient, real variable. |
| $A_j(\mathcal{N}_i)$ | Single-carrier channel coefficient, complex variable, expressed as $A_j(\mathcal{N}_j) = \frac{1}{l}\sum_{i=0}^{l-1} F(T_i(\mathcal{N}_i))$. |
| $\left\|A_j\left(\mathcal{N}_j\right)\right\|^2$ | Single-carrier channel coefficient, real variable, expressed as $\left\|A_j\left(\mathcal{N}_j\right)\right\|^2 = \frac{1}{l}\sum_{i=0}^{l-1}\left\|F\left(T_i\left(\mathcal{N}_i\right)\right)\right\|^2$, real variable. |
| $\mathrm{M}(\cdot)$ | The MGF (Moment-Generating function) function. |
| $P(\cdot)$ | The PDF (Probability Density Function) function. |
| $P_c(\cdot)$ | The common PDF of unordered random variables. |
| $f(\cdot)$ | The CDF (Cumulative Distribution Function) function. |
| $f_c(\cdot)$ | The common CDF of unordered random variables. |
| $R_{err}(x)$ | Error rate function. |
| $\widetilde{R_{err}}$ | Average error rate. |
| $\Lambda_{U_k}$ | Channel selection operator for user $U_k$, $\Lambda_{U_k} = O\Sigma$, where $O$ is the sub-channel ordering operator, while $\Sigma$ is the sum generating operator. |
| $\left\|F\left(T_i^*\left(\mathcal{N}_i\right)\right)\right\|^2$ | Threshold parameter for the sub-channel selection process. If $\frac{1}{l}\left\|F\left(T_i^*\left(\mathcal{N}_i\right)\right)\right\|^2 \leq \frac{1}{l}\left\|F\left(T_i\left(\mathcal{N}_i\right)\right)\right\|^2$ for Gaussian sub-channel $\mathcal{N}_i$, the sub-channel is qualified as "good", otherwise it is put into the "bad" set. |
| $\Lambda_0$ | Sub-channel selection with complete scan at apriori fixed threshold $\left\|F\left(T_i^*\left(\mathcal{N}_i\right)\right)\right\|^2$, which is determined via the optimal Gaussian attack. All sub-channels ($n$) are scanned through, the best $l$ selected. In a modified version of $\Lambda_0$, the threshold is evaluated via the transmittance coefficient of the best available sub-channel as $\frac{1}{l}\left\|F\left(T_i^*\left(\mathcal{N}_i\right)\right)\right\|^2 = \mu\max_l\frac{1}{l}\left\|\left(F\left(T_i\left(\mathcal{N}_i\right)\right)\right)\right\|^2$. |
| $\Lambda$ | Progressive sub-channel selection with an apriori fixed threshold $\left\|F\left(T_i^*\left(\mathcal{N}_i\right)\right)\right\|^2$, at an optimized complexity. The iteration does not require full scan, as the $l$ sub-channels are determined the iteration stops. |
| $\Lambda'$ | Progressive sub-channel selection with apriori fixed threshold $\left|F\left(T_i^*\left(\mathcal{N}_i\right)\right)\right|^2$, at an optimized complexity. If no $l$ sub-channels available at the threshold, the operator selects the remaining sub-channels from the bad channel set (bad: the apriori fixed threshold does not hold for the sub-channel). |
| $\lambda$ | Lagrange multiplier, used to determine the $\left|F\left(T_i^*\left(\mathcal{N}_i\right)\right)\right|^2$ threshold parameter, as $$\lambda = \left|F\left(T_\mathcal{N}^*\right)\right|^2 = \tfrac{1}{l}\sum_{i=0}^{l-1}\left|F\left(T_i^*\left(\mathcal{N}_i\right)\right)\right|^2$$ $$= \tfrac{1}{l}\sum_{i=0}^{l-1}\left|\sum_{k=0}^{l-1} T_k^* e^{\tfrac{-\mathrm{i}2\pi ik}{n}}\right|^2,$$ while $T_\mathcal{N}^*$ is the expected transmittance of the $l$ sub-channels under an optimal Gaussian attack. |
| $\lambda_{\Lambda'}$ | Lagrange multiplier at $\Lambda'$, $$\lambda_{\Lambda'} = \left|F\left(\widehat{T}_\mathcal{N}\right)\right|^2 = \left|F\left(T_i^*\left(\mathcal{N}_i\right)\right)\right|^2 + \varpi,$$ where $\varpi$ is a nonnegative real variable, $$\varpi = \left|F\left(T_i^*\left(\mathcal{N}_i\right)\right)\right|^2$$ $$- \min\left(\left(\left|F\left(T\left(\mathcal{N}_k\right)\right)\right|^2,\ldots,\left|F\left(T\left(\mathcal{N}_{l-1}\right)\right)\right|^2\right)\right),$$ and $\left|F\left(T\left(\mathcal{N}_k\right)\right)\right|^2,\ldots,\left|F\left(T\left(\mathcal{N}_{l-1}\right)\right)\right|^2$ are the corresponding coefficients of the bad set $\mathcal{B}\left(\mathcal{N}_k,\ldots,\mathcal{N}_{l-1}\right)$. |
| $\Gamma_1,\ \Gamma_2$ | Correlated random variables, $\Gamma_1 = \tfrac{1}{l}\sum_{i=0}^{l-2}\left|F\left(T_i\left(\mathcal{N}_i\right)\right)\right|^2$, and $\Gamma_2 = \tfrac{1}{l}\left|F\left(T_{l-1}\left(\mathcal{N}_{l-1}\right)\right)\right|^2$, where the transmittance coefficients are sorted in a descending order $\tfrac{1}{l}\left|F\left(T_0\left(\mathcal{N}_0\right)\right)\right|^2 \geq \tfrac{1}{l}\left|F\left(T_1\left(\mathcal{N}_1\right)\right)\right|^2 \ldots \geq \tfrac{1}{l}\left|F\left(T_{l-1}\left(\mathcal{N}_{l-1}\right)\right)\right|^2$, and $\left|A_j\left(\mathcal{N}_j\right)\right|^2 = \Gamma_1 + \Gamma_2$. |
| $\kappa_{\Lambda_0},\ \kappa_\Lambda,\ \kappa_{\Lambda'}$ | The average number of the iterations (e.g., the number of comparisons of sub-channel transmittance coefficients at a |
| | $\left\vert F\left(T_i^*\left(\mathcal{N}_i\right)\right)\right\vert$ threshold per sub-channels). |
| $p_{err}^{\Lambda_0}\left(A_j\right), p_{err}^{\Lambda}\left(A_j\right), p_{err}^{\Lambda'}\left(A_j\right)$ | Single-carrier level, quadrature error probability at operators $\Lambda_0$, $\Lambda$ and $\Lambda'$. |
| $\mathrm{M}^{\Lambda_0}_{\left\vert A_j(\mathcal{N}_j)\right\vert^2 \cdot \widehat{\mathrm{SNR}}}$ | The MGF function of $\left\vert A_j\left(\mathcal{N}_j\right)\right\vert^2 \cdot \widehat{\mathrm{SNR}}$ at operator $\Lambda_0$. |
| $\mathrm{M}^{\Lambda}_{\left\vert A_j(\mathcal{N}_j)\right\vert^2 \cdot \widehat{\mathrm{SNR}}}(\cdot)$ | The MGF function of $\left\vert A_j\left(\mathcal{N}_j\right)\right\vert^2 \cdot \widehat{\mathrm{SNR}}$ at operator $\Lambda$. |
| $\mathrm{M}^{\Lambda'}_{\left\vert A_j(\mathcal{N}_j)\right\vert^2 \cdot \widehat{\mathrm{SNR}}}$ | The MGF function of $\left\vert A_j\left(\mathcal{N}_j\right)\right\vert^2 \cdot \widehat{\mathrm{SNR}}$ at operator $\Lambda'$. |
| $\eta$ | $\eta$-transform. For a nonnegative random variable $X$ is defined as $\eta_X(\gamma) = \mathbb{E}\left[\frac{1}{1+\gamma X}\right]$, where $\gamma$ is a nonnegative real number, and $0 < \eta_X(\gamma) \leq 1$. |
| $\nu$ | $\nu$-transform (Shannon transform). For a nonnegative random variable $X$ is defined as, $\nu_X(\gamma) = \mathbb{E}\left[\log_2\left(1+\gamma X\right)\right]$, where $\gamma$ is a nonnegative real number. |
| $\nu_{F(\mathbf{T}(\mathcal{N}))F(\mathbf{T}(\mathcal{N}))^\dagger}(\gamma)$ | The $\nu$-transform of $F\left(\mathbf{T}(\mathcal{N})\right)F\left(\mathbf{T}(\mathcal{N})\right)^\dagger$, expressed as $\nu_{F(\mathbf{T}(\mathcal{N}))F(\mathbf{T}(\mathcal{N}))^\dagger}(\gamma) = \nu_{\mathrm{XX}^\dagger}(\chi\gamma_d) + \chi\nu_{\Phi\Phi^\dagger}(\gamma_t) - \chi\frac{\gamma_d\gamma_t}{\gamma}\log_2$, where $\gamma_d$ and $\gamma_t$ are random variables, $\frac{\gamma_d\gamma_t}{\gamma} = 1 - \eta_{\Phi\Phi^\dagger}(\gamma_t)$, $\chi\frac{\gamma_d\gamma_t}{\gamma} = 1 - \eta_{\mathrm{XX}^\dagger}(\chi\gamma_d)$. |
| $\mathcal{D}(x)$ | Distribution function of variable $x$. |
| $\mathcal{P}$ | Probability measure function that returns an event's probability. |
| $f_\chi(x)$ | The Marcenko-Pastur density function, $f_\chi(x) = \left(1 - \frac{1}{\chi}\right)^+ \xi(x) + \frac{\sqrt{(x-a)^+(b-x)^+}}{2\pi\chi x}$, where $\xi(\cdot)$ is a relating random function, $a = \left(1-\sqrt{\chi}\right)^2$ and $b = \left(1+\sqrt{\chi}\right)^2$. |
| $\mathcal{L}$ | Laplace transform. |
| $X \in \mathcal{U}_{[0,1]}, Y \in \mathcal{U}_{[0,1]}$ | Independent, random variables, drawn from a $\mathcal{U}$ uniform distribution on $[0,1]$. |
| $Z \in \mathcal{U}_{[b,1]}$ | Independent random variable, drawn from a $\mathcal{U}$ uniform distribution on $[b,1]$. |
| $\lvert C \rvert$ | Random variable having a distribution of the asymptotic singular value distribution of the singular values of X. The distribution of $\lvert C \rvert^2$ is determined by the asymptotic spectra of $XX^\dagger$. |
| $\lvert \mho \rvert$ | Random variable having a distribution of the asymptotic singular value distribution of the singular values of $\Phi$. The distribution of $\lvert \mho \rvert^2$ is determined by the asymptotic spectra of $\Phi\Phi^\dagger$ |
| $\mathrm{var}(a,b)$ | Variance profile function. |
| $\rho_{F(\mathbf{T}(\mathcal{N}))}(a,b)$ | Channel profile function of $F(\mathbf{T}(\mathcal{N}))$, $\rho_{F(\mathbf{T}(\mathcal{N}))}(a,b):[0,1]^2 \to \mathbb{R}$. |
| $F_a(\cdot), F_b(\cdot)$ | Nonrandom limits. |
| $\chi$ | Nonnegative variable, such that $K/l \to \chi$ holds, where $K$ is the number of users, $l$ is the number of available sub-channels. |
| $\Delta^{(K)}$ | Noise matrix of $K$ users. |
| $\mathcal{M}_k$ | Logical channel of user $U_k$, a set formulated from $m$ sub-carriers such that $\mathcal{M}_k = \left(\mathcal{M}_k^{(0)},...,\mathcal{M}_k^{(l)}\right)^T$, where $\mathcal{M}_k^{(i)}$ identifies the $i$-th logical sub-channel of $U_k$. |
| $P_{sym}(\mathcal{M}_k)$ | Symmetric private classical capacity of the $\mathcal{M}_k$ logical channel of $U_k$, the maximum common rate at which the $K$ users both can reliably transmit private classical information over the $l$ sub-channels of $\mathcal{N}$. |
| $S_{sym}(\mathcal{M}_k)$ | Symmetric secret key rate of the $\mathcal{M}_k$ logical channel of $U_k$, a common rate at which the $K$ users both can reliably transmit private classical information over the $l$ sub-channels of $\mathcal{N}$. |
| $b$ | Ratio of the $k$ actual number of users and the total users $K$, $b = \frac{k}{K}$. |
| $X \circ Z\Phi$ | The random matrix decomposition at arbitrarily allocated sub-channels. Random matrix decomposition of $F(\mathbf{T}(\mathcal{N}))$, where $Z = (\Theta_0,...,\Theta_{K-1})$ is an $l \times K$ random matrix, $\Theta_k = [\odot_{k,0},...,\odot_{k,l-1}]^T$ is user $U_k$'s entry, X is an $l \times K$ random matrix $X = (\mathcal{X}_0,...,\mathcal{X}_{K-1})$, with $\mathcal{X}_i = \left[A_{j,0}^{(k)},...,A_{j,l-1}^{(k)}\right]^T$ for user $U_k$, where $A_j^{(k)} = \frac{1}{m}\sum_m F(T_i(\mathcal{N}_i))$ is the averaged transmittance coefficient of $U_k$ derived by operator $\Lambda_0, \Lambda$ or $\Lambda'$, $m$ is the number of sub-channels of $U_k$, $\circ$ is the Hadamard product operator, while $\Phi$ is an $l \times K$ random matrix, $\Phi = diag(\Phi_0,...,\Phi_{K-1})$. |
| $XU\Phi$ | The random matrix decomposition at identically allocated sub-channels. The $h \times K$ channel matrix $F(\mathbf{T}(\mathcal{N}))$ is $F(\mathbf{T}(\mathcal{N})) = XU\Phi$, where $U = (\mathbf{u}_0,...,\mathbf{u}_{K-1})$ is an $h \times K$ isotropic unitary matrix with arbitrarily distributed random variables with mean $\mu_{i,j} = \mu$ and variance $\sigma_{i,j}^2 = 1/l$, $\mathbf{u}_k = [q_{k,0},...,q_{k,h-1}]^T$, X is an $h \times h$ random matrix $X = diag(\mathcal{X}_0,...,\mathcal{X}_{h-1})$, with $\mathcal{X}_k = \left[A_{j,0}^{(k)},...,A_{j,h-1}^{(k)}\right]^T$, where $A_j^{(k)} = \frac{1}{h}\sum_h F(T_i(\mathcal{N}_i))$ is the averaged transmittance coefficients of $U_k$ in $\mathcal{S}$, such that $\mathcal{X}_i = \mathcal{X}_k$ for $i = 0,...,K-1$, while $\Phi$ is a $K \times K$ random matrix, $\Phi = (\Phi_0,...,\Phi_{K-1})$. |
| $U_{K_{out}}$ | The unitary CVQFT operation, $U_{K_{out}} = \frac{1}{\sqrt{K_{out}}} e^{\frac{-i2\pi ik}{K_{out}}}$, |
| | $i, k = 0, ..., K_{out} - 1$, $K_{out} \times K_{out}$ unitary matrix. |
| $U_{K_{in}}$ | The unitary inverse CVQFT operation, $U_{K_{in}} = \frac{1}{\sqrt{K_{in}}} e^{\frac{i2\pi ik}{K_{in}}}$, $i, k = 0, ..., K_{in} - 1$, $K_{in} \times K_{in}$ unitary matrix. |
| $z \in \mathcal{CN}(0, \sigma_z^2)$ | The variable of a single-carrier Gaussian CV state, $\lvert \varphi_i \rangle \in \mathcal{S}$. Zero-mean, circular symmetric complex Gaussian random variable, $\sigma_z^2 = \mathbb{E}[\lvert z \rvert^2] = 2\sigma_{\omega_0}^2$, with i.i.d. zero mean, Gaussian random quadrature components $x, p \in \mathbb{N}(0, \sigma_{\omega_0}^2)$, where $\sigma_{\omega_0}^2$ is the variance. |
| $\Delta \in \mathcal{CN}(0, \sigma_\Delta^2)$ | The noise variable of the Gaussian channel $\mathcal{N}$, with i.i.d. zero-mean, Gaussian random noise components on the position and momentum quadratures $\Delta_x, \Delta_p \in \mathbb{N}(0, \sigma_\mathcal{N}^2)$, $\sigma_\Delta^2 = \mathbb{E}[\lvert \Delta \rvert^2] = 2\sigma_\mathcal{N}^2$. |
| $d \in \mathcal{CN}(0, \sigma_d^2)$ | The variable of a Gaussian subcarrier CV state, $\lvert \phi_i \rangle \in \mathcal{S}$. Zero-mean, circular symmetric Gaussian random variable, $\sigma_d^2 = \mathbb{E}[\lvert d \rvert^2] = 2\sigma_\omega^2$, with i.i.d. zero mean, Gaussian random quadrature components $x_d, p_d \in \mathbb{N}(0, \sigma_\omega^2)$, where $\sigma_\omega^2$ is the (constant) modulation variance of the Gaussian subcarrier CV state. |
| $F^{-1}(\cdot) = \text{CVQFT}^\dagger(\cdot)$ | The inverse CVQFT transformation, applied by the encoder, continuous-variable unitary operation. |
| $F(\cdot) = \text{CVQFT}(\cdot)$ | The CVQFT transformation, applied by the decoder, continuous-variable unitary operation. |
| $F^{-1}(\cdot) = \text{IFFT}(\cdot)$ | Inverse FFT transform, applied by the encoder. |
| $\sigma_{\omega_0}^2$ | Single-carrier modulation variance. |
| $\sigma_\omega^2 = \frac{1}{l} \sum_l \sigma_{\omega_i}^2$ | Multicarrier modulation variance. Average modulation variance of the $l$ Gaussian sub-channels $\mathcal{N}_i$. |
| $\lvert \phi_i \rangle = \lvert \mathrm{IFFT}(z_{k,i}) \rangle$ $= \lvert F^{-1}(z_{k,i}) \rangle = \lvert d_i \rangle.$ | The $i$-th Gaussian subcarrier CV of user $U_k$, where IFFT stands for the Inverse Fast Fourier Transform, $\lvert \phi_i \rangle \in \mathcal{S}$, $d_i \in \mathcal{CN}(0, \sigma_{d_i}^2)$, $\sigma_{d_i}^2 = \mathbb{E}[\lvert d_i \rvert^2]$, $d_i = x_{d_i} + \mathrm{i} p_{d_i}$, $x_{d_i} \in \mathbb{N}(0, \sigma_{\omega_F}^2)$, $p_{d_i} \in \mathbb{N}(0, \sigma_{\omega_F}^2)$ are i.i.d. zero-mean Gaussian random quadrature components, and $\sigma_{\omega_F}^2$ is the variance of the Fourier transformed Gaussian state. |
| $\lvert \varphi_{k,i} \rangle = \mathrm{CVQFT}(\lvert \phi_i \rangle)$ | The decoded single-carrier CV of user $U_k$ from the subcarrier CV, expressed as $F(\lvert d_i \rangle) = \lvert F(F^{-1}(z_{k,i})) \rangle = \lvert z_{k,i} \rangle$. |
| $\mathcal{N}$ | Gaussian quantum channel. |
| $\mathcal{N}_i, i = 0, \ldots, n-1$ | Gaussian sub-channels. |
| $T(\mathcal{N})$ | Channel transmittance, normalized complex random variable, $T(\mathcal{N}) = \mathrm{Re}\, T(\mathcal{N}) + \mathrm{i}\, \mathrm{Im}\, T(\mathcal{N}) \in \mathcal{C}$. The real part identifies the position quadrature transmission, the imaginary part identifies the transmittance of the position quadrature. |
| $T_i(\mathcal{N}_i)$ | Transmittance coefficient of Gaussian sub-channel $\mathcal{N}_i$, $T_i(\mathcal{N}_i) = \mathrm{Re}(T_i(\mathcal{N}_i)) + \mathrm{i}\, \mathrm{Im}(T_i(\mathcal{N}_i)) \in \mathcal{C}$, quantifies the position and momentum quadrature transmission, with (normalized) real and imaginary parts $0 \leq \mathrm{Re}\, T_i(\mathcal{N}_i) \leq 1/\sqrt{2}$, $0 \leq \mathrm{Im}\, T_i(\mathcal{N}_i) \leq 1/\sqrt{2}$, where $\mathrm{Re}\, T_i(\mathcal{N}_i) = \mathrm{Im}\, T_i(\mathcal{N}_i)$. |
| $T_{Eve}$ | Eve's transmittance, $T_{Eve} = 1 - T(\mathcal{N})$. |
| $T_{Eve,i}$ | Eve's transmittance for the $i$-th subcarrier CV. |
| $\mathbf{z} = \mathbf{x} + \mathrm{i}\mathbf{p} = (z_0, \ldots, z_{d-1})^T$ | A $d$-dimensional, zero-mean, circular symmetric complex random Gaussian vector that models $d$ Gaussian CV input states, $\mathcal{CN}(0, \mathbf{K_z})$, $\mathbf{K_z} = \mathbb{E}[\mathbf{zz}^\dagger]$, where $z_i = x_i + \mathrm{i} p_i$, $\mathbf{x} = (x_0, \ldots, x_{d-1})^T$, $\mathbf{p} = (p_0, \ldots, p_{d-1})^T$, $x_i \in \mathbb{N}(0, \sigma_{\omega_0}^2)$, $p_i \in \mathbb{N}(0, \sigma_{\omega_0}^2)$ i.i.d. zero-mean Gaussian random variables. |
| $\mathbf{d} = F^{-1}(\mathbf{z})$ | An $l$-dimensional, zero-mean, circular symmetric complex random Gaussian vector, $\mathcal{CN}(0, \mathbf{K_d})$, $\mathbf{K_d} = \mathbb{E}[\mathbf{dd}^\dagger]$, $\mathbf{d} = (d_0,...,d_{l-1})^T$, $d_i = x_i + \mathrm{i}p_i$, $x_i, p_i \in \mathbb{N}(0, \sigma^2_{\omega_F})$ are i.i.d. zero-mean Gaussian random variables, $\sigma^2_{\omega_F} = 1/\sigma^2_{\omega_0}$. The $i$-th component is $d_i \in \mathcal{CN}(0, \sigma^2_{d_i})$, $\sigma^2_{d_i} = \mathbb{E}[|d_i|^2]$. |
| $\mathbf{y}_k \in \mathcal{CN}(0, \mathbb{E}[\mathbf{y}_k \mathbf{y}_k^\dagger])$ | A $d$-dimensional zero-mean, circular symmetric complex Gaussian random vector. |
| $y_{k,m}$ | The $m$-th element of the $k$-th user's vector $\mathbf{y}_k$, expressed as $y_{k,m} = \sum_l F(T_i(\mathcal{N}_i))F(d_i) + F(\Delta_i)$. |
| $F(\mathbf{T}(\mathcal{N}))$ | Fourier transform of $\mathbf{T}(\mathcal{N}) = [T_0(\mathcal{N}_0)...,T_{l-1}(\mathcal{N}_{l-1})]^T \in \mathcal{C}^l$, the complex transmittance vector. |
| $F(\Delta)$ | Complex vector, expressed as $F(\Delta) = e^{\frac{-F(\Delta)^T \mathbf{K}_{F(\Delta)} F(\Delta)}{2}}$, with covariance matrix $\mathbf{K}_{F(\Delta)} = \mathbb{E}[F(\Delta)F(\Delta)^\dagger]$. |
| $\mathbf{y}[j]$ | AMQD block, $\mathbf{y}[j] = F(\mathbf{T}(\mathcal{N}))F(\mathbf{d})[j] + F(\Delta)[j]$. |
| $\tau = \|F(\mathbf{d})[j]\|^2$ | An exponentially distributed variable, with density $f(\tau) = (1/2\sigma^{2n}_\omega)e^{-\tau/2\sigma^2_\omega}$, $\mathbb{E}[\tau] \leq n2\sigma^2_\omega$. |
| $T_{Eve,i}$ | Eve's transmittance on the Gaussian sub-channel $\mathcal{N}_i$, $T_{Eve,i} = \mathrm{Re}\, T_{Eve,i} + \mathrm{i}\,\mathrm{Im}\, T_{Eve,i} \in \mathcal{C}$, $0 \leq \mathrm{Re}\, T_{Eve,i} \leq 1/\sqrt{2}$, $0 \leq \mathrm{Im}\, T_{Eve,i} \leq 1/\sqrt{2}$, $0 \leq |T_{Eve,i}|^2 < 1$. |
| $d_i$ | A $d_i$ subcarrier in an AMQD block. |
| $\nu_{\min}$ | The $\min\{\nu_0,...,\nu_{l-1}\}$ minimum of the $\nu_i$ sub-channel coefficients, where $\nu_i = \sigma^2_\mathcal{N}/|F(T_i(\mathcal{N}_i))|^2$ and $\nu_i < \nu_{Eve}$. |
| $\sigma^2_\omega$ | Constant modulation variance, $\sigma^2_\omega = \nu_{Eve} - \nu_{\min}\mathcal{G}(\delta)_{p(x)}$, where $\nu_{Eve} = \frac{1}{\lambda}$, |

$$\lambda = \left|F\left(T_{\mathcal{N}}^{*}\right)\right|^{2} = \tfrac{1}{n}\sum_{i=0}^{n-1}\left|\sum_{k=0}^{n-1} T_{k}^{*} e^{\frac{-\mathrm{i}2\pi ik}{n}}\right|^{2}$$ and $T_{\mathcal{N}}^{*}$ is the expected transmittance of the Gaussian sub-channels under an optimal Gaussian collective attack.

## S.2   Abbreviations

| | |
|---|---|
| AMQD | Adaptive Multicarrier Quadrature Division |
| BS | Beam Splitter |
| CDF | Cumulative Distribution Function |
| CV | Continuous-Variable |
| CVQFT | Continuous-Variable Quantum Fourier Transform |
| CVQKD | Continuous-Variable Quantum Key Distribution |
| DV | Discrete Variable |
| FFT | Fast Fourier Transform |
| IFFT | Inverse Fast Fourier Transform |
| MGF | Moment-Generating Function |
| MQA | Multiuser Quadrature Allocation |
| PDF | Probability Density Function |
| QKD | Quantum Key Distribution |
| SNR | Signal to Noise Ratio |
| SVD | Singular Value Decomposition |